\documentclass{article}
\usepackage{arxiv}

\usepackage[utf8]{inputenc} 
\usepackage[T1]{fontenc}    
\usepackage[hidelinks]{hyperref}     
\usepackage{url}            
\usepackage{booktabs}       
\usepackage{amsfonts}       
\usepackage{nicefrac}       
\usepackage{microtype}      
\usepackage{lipsum}
\usepackage{bm}
\usepackage{graphicx}
\usepackage{subcaption}
\usepackage{amsmath}
\usepackage{placeins}
\graphicspath{ {./images/} }

\title{Nonlocal thermal Willis coupling in laminated conductors}

\author{
  Chunlin Wu \\
  Shanghai Institute of Applied Mathematics \\and Mechanics
  ,Shanghai University\\
  Shanghai, 200044 \\
  \texttt{chunlinwu@shu.edu.cn} \\
\And 
  Gal Shmuel \\ 
  Faculty of Mechanical Engineering, \\
  Technion—Israel Institute of Technology, Israel\\
  \texttt{meshmuel@technion.ac.il}
  \And
  Huiming Yin\thanks{Corresponding Author} \\
  Department of Civil Engineering and \\Engineering Mechanics,
  Columbia University\\
  New York, NY, 10027 \\
  \texttt{yin@civil.columbia.edu} \\
}

\begin{document}
\maketitle
\begin{abstract}
Building on Willis' homogenization framework, recent work has revealed that heterogeneous conductors exhibit macroscopic thermal bianisotropy, in which the macroscopic heat flux and entropy are nonlocally coupled to both temperature and temperature gradient. Existing numerical examples, however, are limited to the subwavelength regime. Here, we provide the first explicit demonstration of this spatial nonlocality by computing the effective kernels of a periodic laminate using three independent homogenization methods. The three approaches yield consistent nonlocal cross-coupling terms, clarifying the roles of spatial asymmetry and averaging choice. We also calculate the corresponding thermal impedance and show that it is direction-dependent, highlighting a physical signature of thermal bianisotropy relevant to thermal metamaterials.
\end{abstract}


\section{Introduction}

Metamaterials are artificial composites engineered to exhibit extraordinarily effective properties across various physical responses \cite{Kadic2013RPP,Kadi2019nrp,srivastava2015elastic,Christensen2015MRCComunications}. In particular, thermal metamaterials are designed to redirect heat flow in ways that natural materials cannot \cite{Schittny2013,Su2023,li2021natrevmat, Zhang2023natphysrev}. Representative examples include thermal cloaks \cite{Yang2020}, concentrators \cite{Shen2016} and camouflage devices \cite{Zhou2018}.

A rigorous route to derive the effective properties of composites and metamaterials is provided by homogenization theories, which use an averaging process to replace a heterogeneous medium by an equivalent uniform description \cite{milton2002theory,Simovski2009bc,tart00book}. Classical homogenization methods are commonly based on asymptotic expansions and scale-separation assumptions \cite{Kaminski2003,Kaminski2003a,Xu2014,Matine2015,Haymes2017,Antonakakis2013PRSA,Nassar2016}. These methods are powerful in the long-wavelength or low-frequency regime, but generally do not capture the full nonlocal response of heterogeneous media. Willis developed an exact homogenization framework for elastodynamics that applies beyond these regimes and reveals nonlocal effective properties that appear only at the macroscopic scale \cite{Willis1980,willis1981variational,Willis1985IJSS,Willis1997,Willis2009,Willis2011PRSA}. The corresponding cross-couplings, now known as Willis couplings, constitute additional knobs for wave manipulation and have therefore attracted  interest in the metamaterials community \cite{Sieck2017prb,quan2018prl,Melnikov2019nc,Liu2019prx}.

By extending Willis' framework to piezoelectric composites, Pernas-Salom\'on and Shmuel derived analogous electro-momentum couplings \cite{PernasSalomon2020,pernassalomn2020prapplied}. More recently, Shmuel and Willis adapted this framework to heat conduction \cite{Shmuel2025,Gal2025a}, demonstrating nonlocal thermal bianisotropic response by a heterogeneous conductor. They showed that, macroscopically, the heat flux is nonlocally coupled not only to the temperature gradient, but also to the temperature itself, while the entropy or heat storage is nonlocally coupled not only to the temperature, but also to its gradient. This notion of thermal Willis coupling was introduced earlier by Torrent et al. \cite{Torrent2018} and Xu et al. \cite{Xu2022} using media whose thermal properties are modulated in both space and time. By contrast, the thermal bianisotropic framework of Shmuel and Willis shows that time modulation is not necessary: spatial asymmetry alone is sufficient.

While the exact homogenization theory predicting nonlocal thermal bianisotropy has already been established \cite{Shmuel2025,Gal2025a}, explicit calculations of the corresponding effective kernels remain limited to the subwavelength regime (in elastodynamics, these constitutive equations are known as the Milton-Briane-Willis equations \cite{milton06cloaking,Milton2007njp,Milton2020II,Milton2020IV}). Here, we provide the first explicit numerical demonstration of this nonlocality in spatially asymmetric composites by calculating the effective kernels of a periodic laminate.

The effective properties are not uniquely defined unless a sufficient set of independent driving sources is admitted, as first recognized by Fietz and Shvets \cite{FIETZ2010pysicaaB} and Willis \cite{Willis2011PRSA}, and discussed later in several works \cite{alu2011prb,Nassar2015,Sieck2017prb,pernassalomn2020prapplied,Gal2025a}. At a basic level, this nonuniqueness has two origins. In the absence of volumetric sources, the effective flux-like fields are dependent because of the balance law; in the absence of residual kinematic fields, the effective kinematic fields are dependent because they derive from the same potential. Here, similarly to the formulation of Willis \cite{Willis2009}, we account for a volumetric heat source but do not introduce an independent residual temperature-gradient field. The resulting effective relations should therefore be understood as one admissible representative of the thermal Willis response. In this representative, the thermal Willis couplings vanish for reflection-symmetric microstructures, even at finite wavelength. By contrast, when a residual field is admitted, certain nonlocal effects are captured by the cross-couplings rather than being absorbed into the direct effective kernels \cite{Sieck2017prb,PernasSalomon2020,Muhafra2022}.

We consider two conducting phases governed by Fourier's law and a localized heat-capacity inclusion. We regard different translations of the unit cell as different realizations of a random medium, so that ensemble averaging provides the macroscopic fields \cite{hashin1983analysis,Willis2011PRSA}. The position of the localized heat-capacity inclusion breaks the reflection symmetry of the unit cell and controls the magnitude and phase of the emergent coupling terms. By varying the inclusion position and the associated asymmetry, we quantify how the microstructure controls the thermal bianisotropic response.

We compute this representative  effective set using three different methods. First, we develop a polarization formulation based on an infinite-domain Green's function and equivalent eigen-fields, in the spirit of Eshelby's equivalent inclusion method. The motivation for this approach is that the exact Green's function of a specific problem is often unknown for more general microstructures; the polarization formulation instead uses an infinite-domain comparison Green's function and represents material mismatch through equivalent eigen-fields. Second, we derive the nonlocal kernels directly from the Green's function of the periodic laminate, following the  construction used by Willis for laminates in elastodynamics \cite{Willis2009}. Third, we compare these results with a modified boundary-retrieval method.

This comparison allows us to examine not only the emergence of the thermal bianisotropic terms, but also the role of the averaging procedure used to define the macroscopic fields. The three homogenization methods yield consistent nonlocal cross-coupling kernels and recover the expected Willis-type correlation between them. We further show that this agreement depends on the averaging procedure: the de-phased average isolates the relevant Bloch component and produces a consistent constitutive description, whereas the conventional spatial average can obscure the correlation between the coupling kernels.

Finally, we connect the effective kernels to a measurable physical signature by calculating the corresponding thermal impedance. The bianisotropic coupling renders this impedance direction dependent, in analogy with Willis and bianisotropic wave media. Thus, the nonlocal kernels are not merely formal quantities: they determine the phase and magnitude of the macroscopic thermal response, and provide a route for identifying thermal bianisotropy in asymmetric thermal metamaterials.

The paper is structured as follows. Section~2 defines the laminate, the averaging procedures, and the effective thermal bianisotropic constitutive law. Section~3 develops the polarization formulation based on the infinite-domain Green's function. Section~4 derives the exact nonlocal kernels using the periodic Green's function and ensemble averaging over translations of the unit cell. Section~5 presents numerical examples, compares the three homogenization routes, and analyzes the effects of spatial asymmetry, microstructure, averaging, and thermal impedance. Section~6 summarizes the main conclusions.

\section{Problem statement}
\begin{figure}
    \centering
    \includegraphics[width=1\linewidth]{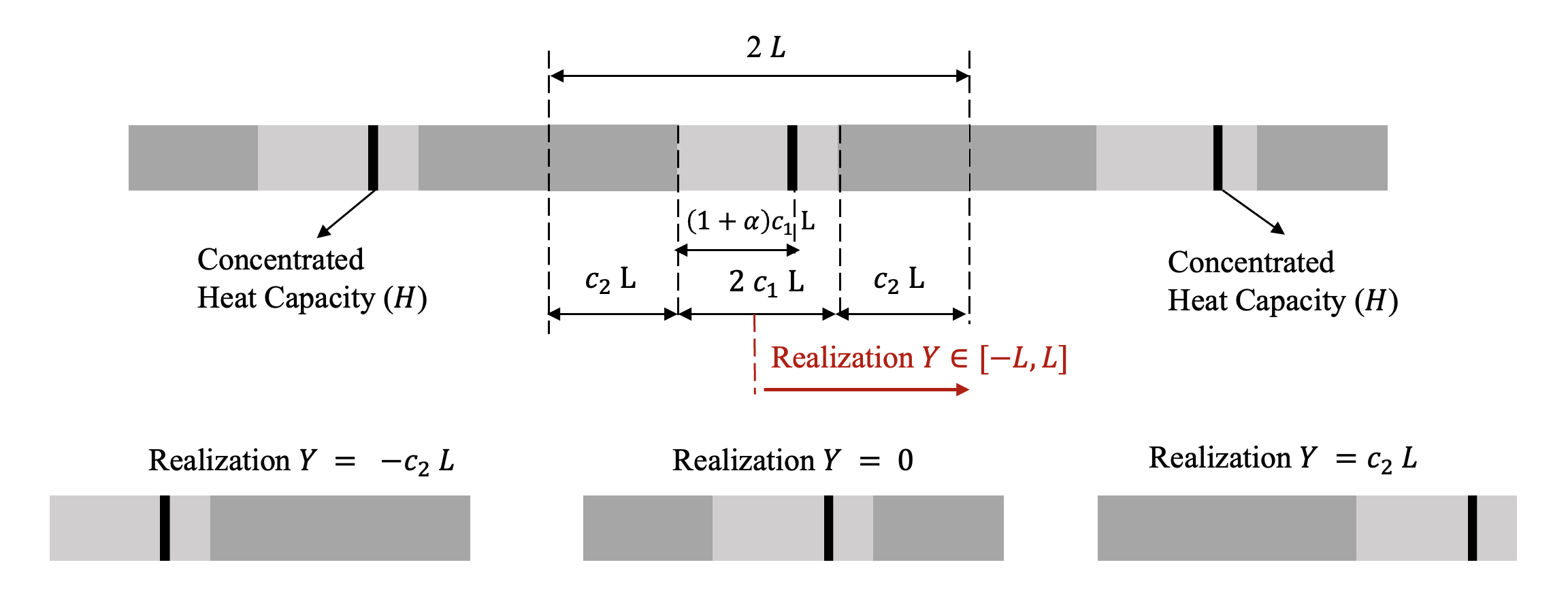}
    \caption{Schematic plot of a periodically infinite laminate composite, and the length of the unit cell is $2 L$. The unit cell is composed of two material phases, the dark gray phase occupies $2 c_2 L$ and the light gray possesses $2 c_1 L$ ($c_1 = 1 - c_2$), while one concentrated heat capacity $H$ (at $p_{Y=0} = \alpha c_1 L$ when $Y = 0$). The realization variable $Y$ ($Y \in [-L, L]$) determines the starting and ending point of the unit cell, and three cases $Y = -c_2 L, 0, c_2 L$ are plotted.}
    \label{fig:problem}
\end{figure}

Fig. \ref{fig:problem} schematically illustrates a periodic one-dimensional (1D) laminate composite, whose unit cell is defined with the length $2 L$. Note that the method and formulae (except the specific Green's function) are general for composites. Without the loss of any generality, we follow Willis' work (in elastodynamics) \cite{Willis2009} to investigate a representative case of the bimaterial laminate. The bimaterial laminate is composed of two isotropic material phases, (i) phase 1 (light gray) occupies $2 c_1 L$; (ii) phase 2 (dark gray) occupies $2 c_2 L$ ($c_1 + c_2 = 1$). In general, each isotropic material phase exhibits different material properties, and $K^{i}, C^{i}$  ($i=1,2$) refer to the thermal conductivity and heat capacity of the material phase $1, 2$, respectively. Based on Shmuel and Willis' observation using the boundary retrieval method (BRM) \cite{Gal2025a}, when the unit cell microstructure is symmetric, the coupling terms vanish, which will be confirmed subsequently in this paper using two other methods, and was also mentioned in elastodynamics with the ensemble average  \cite{Willis2009}. Therefore, adding a concentrated heat capacity of a black line can break the symmetry condition, which yields nonzero coupling terms. More details are provided in Section 5 with numerical simulation. We add the concentrated heat capacity ($H = \alpha C^1 L$) in the unit cell at point $p$. 

For a laminate composite, the effective material properties (by the unit cell) are determined by the starting and ending point of the unit cell, or they can be characterized by a realization variable (coordinate), i.e., $Y \in [-L, L]$. With the realization variable, material properties can be expressed as shifting from the case $Y = 0$, i.e., $K_Y(x) = K_{Y=0}(x-Y)$ and $C_Y(x) = C_{Y=0}(x-Y)$. Specifically, the material properties when $Y = 0$ can be expressed as, 

\begin{equation}
    K_{Y=0}(x) = \begin{cases}
        K^2 & x \in[-L, -c_1 L) \\ 
        K^1 & x \in [-c_1 L, c_1 L) \\
        K^2 & x \in [c_1 L, L]
    \end{cases} \quad \text{and} \quad C_{Y=0}(x) = \begin{cases} C^2 & x \in[-L, -c_1 L) \\ 
        C^1 & x \in [-c_1 L, c_1 L) \\
        C^2 & x \in [c_1 L, L] \end{cases} \quad \text{and} \quad p_{Y=0} = \alpha c_1 L
        \label{eq:mat_phase}
\end{equation}

Two assumptions are stated: (i) the material phases 1 and 2 are fully bonded, whose interface admits continuous temperature $T$ and heat flux $q$; and (ii) the unit cell is taken from the infinite periodic composite, which is subjected to the Bloch-form boundary conditions as $T(-L) = T(L) \exp [-i 2 \zeta L]$ and $q(-L) = q(L) \exp[-i 2 \zeta L]$, in which $\zeta$ represents the thermal wavenumber. 

The heat transfer process in the one-dimensional solid can be governed by the heat equation given the realization variable $Y$, 

\begin{equation}
    C_Y(x) \overline{T}_{,t}(x, t) - \overline{Q}(x, t) = \left[K_Y(x) \overline{T}_{,x}(x, t)\right]_{,x}
    \label{eq:heat_eqn_trans}
\end{equation}
where $\overline{T}(x, t)$ represents the spatiotemporal distribution of temperature; $\overline{Q}(x, t)$ is the spatiotemporal distributed heat source; $\overline{T}_{,t}$ and $\overline{T}_{,x}$ refers to time and spatial partial derivatives of the temperature field. When thermal quantities are sinusoidal functions of time, the temperature can be expressed as $\overline{T}(x, t) = T_R + Re(T(x) \exp[-i \omega t])$, in which $\omega$ is the circular frequency and $T_R$ the reference of the mean temperature. In such a case, the transient heat equation reduces to the time-harmonic case \cite{Wu2025a} by dropping the term of $\exp[-i \omega t]$: 

\begin{equation}
    \left[ K_Y(x) T_{,x}(x) \right]_{,x} + i \omega C_Y(x) T(x) + Q(x) = 0
    \label{eq:heat_eqn_harm}
\end{equation}

If a composite material can reproduce the same macroscopic material behavior as the pure material on average, a correspondence may be established between the effective properties of the homogenized composite and those of the pure material. However, such correspondence may not always exist, particularly considering thermal Willis coupling \cite{Xu2022}.

In ensemble average, $q_Y(x)$ and $F_Y(x)$ for realization $Y$ represent the spatial part of heat flux by local temperature gradient and sensible heat by local temperature change in Eq. (\ref{eq:heat_eqn_harm}) as: 

\begin{equation}
    q_Y(x) = -K_Y(x) T_{Y,x}(x) \quad \text{and} \quad F_Y(x) = C_Y(x) T_Y(x)
    \label{eq:cons_local}
\end{equation}
where $T_{Y}$ and $T_{Y,x}$ are temperature and temperature gradients for realization $Y$. Note that although the local thermal responses are of significance for thermal management and design, the effective thermal properties characterize the overall thermal performances of the composites. 

To define the effective thermal properties, it is natural to employ averaged thermal quantities. For instance, the de-phased (spatial) average \cite{NematNasser2011} and ensemble average \cite{Willis1997} can be applied. Taking the unit cell with the length $2 L$ and the realization variable $Y$, the spatial average of a function $T_Y(x)$ can be defined as: 

\begin{equation}
    \langle T \rangle_Y = \frac{1}{2L} \int_{-L}^{L} T_Y(x) \thinspace f(x) \thinspace dx
    \label{eq:sp_avg}
\end{equation}
where $\langle. \rangle_Y$ refers to the volume average of the function with the realization $Y$; $f(x)$ is the weight function for the spatial average. For example, when $f(x) = 1$, it becomes the spatial average; when $f(x) = \exp[-i \zeta x]$, the weighted average becomes the de-phased average \cite{NematNasser2011}, which helps to filter the periodic part from the temperature variations. Note that although the de-phased average is a common choice in the literature, other weight functions can also be employed; some discrepancies are observed and discussed in Section 5. 

Since the thermal quantities vary with the realization $Y$, the ensemble average is defined to evaluate its expectation as, 

\begin{equation}
 \langle T \rangle (x) = \int_{-L}^{L} T_Y(x) \thinspace w_Y(x) \thinspace \mathcal{P}(Y) \thinspace dY 
    \label{eq:en_avg}
\end{equation}
where $\langle.\rangle$ is the ensemble average with $\mathcal{P}(Y)$ representing the probability distribution of the realization variable $Y$; 
and $w_Y(x)$ is a realization dependent weight function \cite{Willis2009}. In this paper, we assume the uniform probability distribution that $\mathcal{P}(Y) \equiv \frac{1}{2L}$. Compared with the (weighted) spatial average in Eq. \eqref{eq:sp_avg}, the ensemble-average in Eq. \eqref{eq:en_avg} depends on the position $x$. Therefore, Willis' effective constitutive relations are non-local, which can be interpreted with the convolution operations, see explanations in Eqs. (1-5) in \cite{Willis2012}. In analogy with elastodynamics \cite{Willis2012,Srivastava2015}, the effective thermal properties cannot be sufficiently described by the effective Fourier law, i.e., the correlation between averaged temperature gradient and heat flux, because such ``effective'' properties may not satisfy the governing equation on average. In contrast, the effective constitutive relations are bi-anisotropic \cite{Gal2025a} as follows:  

\begin{equation}
     \begin{bmatrix} \langle q \rangle \\ \langle F \rangle \end{bmatrix} = \begin{bmatrix} - K^{\text{eff}} & \chi^{\text{eff}} \\ \xi^{\text{eff}} & C^{\text{eff}} \end{bmatrix} * \begin{bmatrix} \langle T_{,x} \rangle \\ \langle T \rangle \end{bmatrix}
    \label{eq:willis_law_NL}
\end{equation}
where the symbol $(*)$ refers to spatial convolution, i.e., $a * b = \int_{-L}^{L} a(x - x') b(x') dx'$; $K^{\text{eff}}, \chi^{\text{eff}}, \xi^{\text{eff}}$ and $C^{\text{eff}}$ are components of the non-local effective constitutive relations, which are shift-invariants. Because the layered composites are assumed to be in the Bloch-form with the periodic boundary conditions, Eq. (\ref{eq:willis_law_NL}) can be written in terms of the Fourier space as follows: 

\begin{equation}
     \begin{bmatrix} \langle \tilde{q} \rangle \\ \langle \tilde{F} \rangle \end{bmatrix} = \begin{bmatrix} -\tilde{K}^{\text{eff}} & \tilde{\chi}^{\text{eff}} \\ \tilde{\xi}^{\text{eff}} & \tilde{C}^{\text{eff}} \end{bmatrix} \thinspace \begin{bmatrix} \langle \tilde{T}_{,x} \rangle \\ \langle \tilde{T} \rangle \end{bmatrix}
     \label{eq:willis_law}
\end{equation}
where $\tilde{\langle.\rangle} = \int_{-\infty}^{\infty} \langle.\rangle(x) \exp[i \zeta x] \thinspace dx$ is the Fourier transform of quantities. Note that the effective constitutive relations in Eq. (\ref{eq:willis_law}) depend on the thermal wavenumber $\zeta$ and frequency $\omega$. Note that as Eq. (\ref{eq:willis_law}) employs the sensible heat storage versus the temperature change, which can be modified as the sensible heat change rate and temperature change rate by adding the factor $-i \omega$,

\begin{equation}
    \begin{bmatrix} \langle \tilde{q} \rangle \\ -i \omega \langle \tilde{F} \rangle \end{bmatrix} = \begin{bmatrix} -\tilde{K}^{\text{eff}} & \frac{\tilde{\chi}^{\text{eff}}}{-i\omega} \\ -i \omega \tilde{\xi}^{\text{eff}} & \tilde{C}^{\text{eff}} \end{bmatrix} \thinspace \begin{bmatrix} \langle \tilde{T}_{,x} \rangle \\ -i \omega \langle \tilde{T} \rangle \end{bmatrix}
     \label{eq:willis_law_rate}
\end{equation}
where the cross-coupling terms are altered accordingly for consistency with the previous work \cite{Gal2025a}, in which the entropy increment in one cycle was used in place of the sensible heat for small temperature cycling and the factor $-i \omega$ was eliminated with a simpler form. However, this paper focuses on effective properties defined in Eq. (\ref{eq:willis_law}). The Willis thermal homogenization in Eq. (\ref{eq:willis_law}) shows that the macroscopic (global) thermal response is governed by four non-local effective kernels, which correlate the averaged temperature gradient/temperature with the averaged heat flux/sensible heat storage. Therefore, the primary goal of this paper is to determine such kernels that satisfy the heat equation on average. In what follows, we first propose the extraction framework based on Willis' polarization approach \cite{Willis1980} with the Green's function for the infinite domain. Then we derive the effective kernels through the source-driven concept based on microstructural-specific Green's function \cite{Willis2009} and the convolution properties of the Fourier space. 

\section{Effective properties by Eshelby's equivalent inclusion method}
In the pioneering studies \cite{Willis1980, Willis1997, NematNasser2013}, the authors proposed to use the comparison medium and polarization fields to simulate microscopic solutions in elastodynamics, such as the eigenstress and eigenmomentum. It should be noted that when the matrix is selected as the comparison medium, the polarization method reduces to Eshelby's equivalent inclusion method (EIM). Recently, Wu et al. \cite{Wu2025a} have proposed to use eigen-temperature-gradient (ETG) and eigen-heat-source (EHS) to simulate material mismatches in thermal conductivity and heat capacity, respectively. In the following, this subsection proposes a general approach to detect effective properties with the EIM and the infinite Green's function. 

\subsection{Formulation}

The Green's function for harmonic heat transfer in an infinite one-dimensional domain can be written as, 

\begin{equation}
    G^H(x, x') = \frac{i}{2 K^c \gamma} \exp [i \gamma r], \quad \gamma = \frac{\sqrt{2}}{2} (1 + i) \sqrt{\frac{\omega C^c}{K^c}}, \quad r = |x - x'|
    \label{eq:Gfunc_inf}
\end{equation}
where $K^c$ and $C^c$ refer to the thermal conductivity and heat capacity of the comparison medium, and $G^{H}$ is the Green's function for the comparison medium. For consistency, the comparison medium is retained here, while the numerical examples employ the material phase $1$ instead. Using the ETG and EHS \cite{Wu2025a}, the heat equation can be expressed as, 

\begin{equation}
    K^c \frac{\partial}{\partial x} \left[ T_{Y,x}(x) - u_Y^*(x) \right] + i \omega C^c T_Y(x) + Q_Y^*(x) + Q(x) = 0
    \label{eq:heat_equiv}
\end{equation}
where $u^*_Y(x)$ and $Q_Y^*(x)$ are ETG and EHS with the realization $Y$, respectively. Since the Green's function $G^H$ satisfies the governing equation, $K^c G^H_{,xx}(x, x') + i \omega C^c G^H(x, x')= -\delta(x - x')$, the temperature at any interior field can be obtained with the filtering effect of the Dirac Delta function \cite{Wu2025-ijss}:  

\begin{equation}
\begin{split}
    T_Y(x) & = 
    \int_{-L}^{L} \delta(x - x') T_Y(x') \thinspace dx' = -\int_{-L}^{L} \left[ K^c G_{,x'x'}^H(x,x') + i \omega C^c G^H(x, x') \right] \thinspace T_Y(x') \thinspace dx' \\
    &=\underbrace{-\left[ K^c G^H_{,x'}(x, L) T_Y(L) - K^c G^H_{,x'}(x, -L) T_Y(-L) \right] - \left[ G^H(x, L) q_Y(L)  - G^H(x, -L) q_Y(-L) \right]}_{T_Y^B(x)} \\ 
    & \qquad + \underbrace{\int_{-L}^{L} -K^c G^H_{,x}(x, x')  u_Y^*(x') \thinspace dx'}_{T_Y^E(x)} + \underbrace{\int_{-L}^{L} G^H(x, x')  Q_Y^*(x') \thinspace dx'}_{T_Y^Q(x)} + \underbrace{\int_{-L}^{L} G^H(x, x') Q(x') \thinspace dx'}_{T_Y^M(x)}
\end{split}
    \label{eq:bie_1d}
\end{equation}
where $K^c T_{Y,x'x'} = K^c u^*_{Y,x'} - i \omega C^c T_Y - Q^*_Y - Q$ from Eq. (\ref{eq:heat_equiv}) is applied; the temperature and its gradient can be decomposed into four parts as denoted by the superscripts $B$ (boundary), $E$ (ETG), $Q$ (EHS), and $M$ (internal heat source), respectively. Following our recent work \cite{Wu2025a}, two eigen-fields can be determined by the Eshelby's equivalent conditions: 

\begin{equation}
    u_Y^*(x) = \frac{K^c - K_Y(x)}{K^c} T_{Y,x}(x), \quad Q_Y^* = i \omega \left[  C_Y(x) - C^c \right] T_Y(x)
    \label{eq:equiv}
\end{equation}

Although the determination of two eigen-fields is straightforward with Eshelby's EIM, note that the concentrated heat capacity requires proper handling. Since the point heat capacity is $H \delta (x - p_Y)$, the equivalent condition for EHS at the point should only be interpreted in the sense of integration. Specifically, the EHS at $p_Y$ should be written as $Q_Y^{*}(p_Y) = R_Y^* \delta(x - p_Y)$. For instance, if $C^1$ is used as the heat capacity of the comparison medium, as the point heat capacity is also located in the material phase $1$, Eshelby's equivalent inclusion condition is written as:

\begin{equation}
    -i\omega C^1 T_Y(x) - R_Y^*\delta(x - p_Y) = -i \omega [C^1 + H \delta (x - p_Y)] T_Y(x) 
\end{equation}
which leads to 
\begin{equation}
    R_Y^* = i \omega H T_Y(p_Y)
\end{equation}

To numerically evaluate the responses of the composites, the composite domain is divided into $N$ equally small line elements (length $\frac{2L}{N}$) with constant eigen-fields. Note that other shape functions, such as linear and quadratic eigen-field distributions, can be applied in the same procedure with the versatility of Green's function \cite{wu2021elastic}. For simplicity, this paper  only considers constant eigen-fields over elements. The detailed verification and convergence results are elaborated in Section 4.2 of the Supplemental Material. Hence, eigen-fields can be expressed in terms of the Heaviside function $\Theta(.)$ of each line element as follows:

\begin{equation}
\begin{aligned}
    u_Y^*(x) &= \sum_{I=0}^{N} u_Y^{I*} \thinspace \Theta\left[ \left(\frac{L}{N}\right)^2 -(x - x^{IC})^2 \right]\\
    \quad Q_Y^*(x) &= \sum_{I=0}^{N} Q_Y^{I*} \thinspace \Theta\left[ \left(\frac{L}{N}\right)^2 -(x - x^{IC})^2 \right] + R_Y^* \delta(x - p_Y) 
\end{aligned}
    \label{eq:eigen_field}
\end{equation}
where $u_Y^{I*}$ and $Q_Y^{I*}$ refers to the constant ETG and EHS within the $I^{th}$ line element; $x^{IC}$ refers to the center of the $I^{th}$ line element; and the property $\Theta\left[ \left(\frac{L}{N}\right)^2 -(x - x^{IC})^2 \right] = \Theta(\frac{L}{N} - (x - x^{IC}) ) \Theta( \frac{L}{N} + (x - x^{IC}) )$ is applied. The disturbances by eigen-fields can be derived through domain integrals of Green's function, 

\begin{equation}
\begin{aligned}
    T_Y^Q(x) = 
    G^H(x, p_Y) R^* + \sum_{I=0}^{N} Q_Y^{I*} \int_{x^{IC} - \frac{L}{N}}^{x^{IC} + \frac{L}{N}} G^H(x, x') \thinspace dx' = G^H(x, p_Y) R_Y^* + \sum_{I=0}^{N} L^I(x) Q_Y^{I*} \\ 
    T_Y^E(x) = 
    \sum_{I=0}^{N} u_Y^{I*} \int_{x^{IC} - \frac{L}{N}}^{x^{IC} + \frac{L}{N}} -K^c G_{,x}^H(x, x') \thinspace dx' = \sum_{I=0}^{N} -K^c L^I_{Y,x}(x) u^{I*} = \sum_{I=0}^{N} S^I(x) u_Y^{I*}
\end{aligned}
    \label{eq:disturbances}
\end{equation}
where $L^{I}(x)$ is the domain integral of Green's function over the $I^{th}$ line element, which is known as Eshelby's tensor in 3D cases \cite{mura2013micromechanics};  and $S^I(x) = -K^c L_{,x}^I(x)$. Although the 3D Eshelby's tensors reduce to scalars in 1D cases, we use the same terminology, and provide the closed-form Eshelby's tensors for a 1D line element starting at $a$ and ending at $b$ ($b > a$) as below: 

\begin{equation}
    L(x) = \int_{a}^{b} \frac{i}{2 K^c \gamma} \exp [i \gamma r] \thinspace dx' = \frac{-1}{2 K^c \gamma^2} \begin{cases} 
    \exp[i \gamma (a - x)] - \exp[i \gamma (b - x)] & x \in (-\infty, a) \\ 
    2 - \exp[i \gamma (x - a)] - \exp[i \gamma (b - x)] & x \in [a, b] \\
    \exp[i \gamma (x - b)] - \exp[i \gamma (x - a)] & x \in (b, +\infty)
    \end{cases}
\end{equation}
and its first order derivative, 
\begin{equation}
    L_{,x}(x) = \frac{d}{dx} \int_{a}^{b} \frac{i}{2 K^c \gamma} \exp [i \gamma r] \thinspace dx' = \frac{-i}{2 K^c \gamma} \begin{cases} 
    \exp[i \gamma (b-x)]  - \exp[i \gamma (a-x)] & x \in (-\infty, a) \\ 
    \exp[i \gamma (b - x)] - \exp[i \gamma (x - a)]  & x \in [a, b] \\
    \exp[i \gamma (x - b)] - \exp[i \gamma (x - a)]  & x \in (b, +\infty)
    \end{cases}
\end{equation}

A global system of linear equation can be constructed with the unknown vector including boundary responses and eigen-fields. The detail procedure is elaborated in the Supplemental Material 4.1. This paper focuses on the Bloch-form boundary condition for an exact solution as Willis proposed \cite{Willis2009}. Using the Bloch-form boundary condition, the unknowns on the boundary, including temperature and heat flux on the left/right ends, are represented by two independent unknowns. Without loss of generality, the temperature and flux at the left end are selected, $T^L_Y, q^{L}_Y$. Formally, the global linear equation system can be reorganized by combining dependent boundary unknowns as, 

\begin{equation}
    \mathcal{F}_Y \mathcal{S}_Y = \mathcal{M}_Y \mathcal{Q}
    \label{eq:global_form}
\end{equation}
where the subscript $Y$ stands for the realization; $\mathcal{F}_Y$ represents the combined coefficient matrix; $\mathcal{S}_Y$ refers to the list of unknown variables, including boundary temperature/heat flux and eigen-fields; and $\mathcal{M}_Y$ and $\mathcal{Q}$ introduces the effects of internal heat source. More details on rearrangement of global coefficient matrices are provided in Section 4.1 of the Supplemental Materials. Therefore, the boundary responses and eigen-fields can be obtained from Eq. (\ref{eq:global_form}). Because the coefficient matrix $\mathcal{F}_{Y}$ is full-rank, the solution vector and the internal heat source are correlated as, 

\begin{equation}
    \mathcal{S}_Y = \left(\mathcal{F}_Y \right)^{-1} \mathcal{M}_Y \mathcal{Q}
    \label{eq:sol_Q}
\end{equation}
Eq. (\ref{eq:sol_Q}) shows that all boundary responses and eigen-fields are linearly dependent on the prescribed source term, which forms the basis for the source-driven scheme to extract effective properties. 

\subsection{Extraction of effective properties}
Once the boundary responses and eigen-fields are solved, all field thermal quantities at sampling points can be calculated by linear operations. For instance, one can calculate temperature, temperature gradients, heat flux, and sensible heat, as well as their unweighted/weighted averages, in post-processing using domain integrals. Finally, we can extract effective constitutive relations. Using Eq. (\ref{eq:bie_1d}), the temperature and temperature gradients at any interior points, i.e., the $p^{th}$- sampling point $x_p$, can be obtained as, 

\begin{equation}
    \begin{bmatrix} \vdots \\ T_{Y,x}(x_p) \\ T_Y(x_p) \\ \vdots \end{bmatrix} = \underbrace{\begin{bmatrix} 
    \vdots & \vdots & \vdots & \vdots & \vdots & \vdots & \vdots \\
                    -\mathcal{H}'f\thinspace w_Y & \mathcal{G}'f\thinspace w_Y & \ldots & L^{I\prime}f\thinspace w_Y & G^{H\prime}f\thinspace w_Y & S^{I\prime}f\thinspace w_Y  & \ldots \\ 
                    -\mathcal{H}f\thinspace w_Y & \mathcal{G}f\thinspace w_Y & \ldots & L^I f\thinspace w_Y & G^H f \thinspace w_Y &  S^I f\thinspace w_Y  & \ldots \\ 
                    \vdots & \vdots & \vdots & \vdots & \vdots & \vdots & \vdots 
    \end{bmatrix}}_{\mathcal{A}_Y}  \underbrace{\begin{bmatrix} T_Y \\ q_Y \\ \vdots \\ Q_Y^{I*} \\ R_Y^* \\ u_Y^{I*} \\ \vdots \end{bmatrix}}_{\mathcal{S}_Y} + \underbrace{\begin{bmatrix} \vdots \\ L' f\thinspace w_Y \\ L f\thinspace w_Y \\ \vdots \end{bmatrix}}_{\mathcal{E}^A_Y} \underbrace{\begin{bmatrix} Q \end{bmatrix}}_{\mathcal{Q}}
    \label{eq:temp_eqn}
\end{equation}
and the heat flux and sensible heat can be acquired by $q_Y(x_p) = -K_Y T_{Y,x}(x_p)$ and $F_Y(x_p) = C_Y T_Y(x_p)$ as: 
\begin{equation}
\begin{split}
    \begin{bmatrix} \vdots \\ q_Y(x_p) \\ F_Y(x_p) \\ \vdots \end{bmatrix} & = \underbrace{\begin{bmatrix} 
    \vdots & \vdots & \vdots & \vdots & \vdots & \vdots & \vdots \\
    K_Y\mathcal{H}'f \thinspace w_Y & -K_Y\mathcal{G}' f \thinspace w_Y & \ldots & -K_Y L^{I \prime} f \thinspace w_Y & -K_Y G^{H \prime} f\thinspace w_Y &  -K_Y S^{I \prime} f\thinspace w_Y  & \ldots \\ 
                    -C_Y\mathcal{H} f \thinspace w_Y & C_Y \mathcal{G} f \thinspace w_Y & \ldots & C_Y L^{I} f\thinspace w_Y & C_Y G^{H } f \thinspace w_Y & C_Y S^{I}f \thinspace w_Y & \ldots \\ 
                    \vdots & \vdots & \vdots & \vdots & \vdots & \vdots & \vdots 
    \end{bmatrix}}_{\mathcal{B}_Y} \\ 
    & \quad \cdot \underbrace{\begin{bmatrix} T_Y \\ q_Y \\ \vdots \\ Q_Y^{I*} \\ R_Y^* \\ u_Y^{I*} \\ \vdots \end{bmatrix}}_{\mathcal{S}_Y} + \underbrace{\begin{bmatrix} \vdots \\ -K_Y L' f\thinspace w_Y \\ C_Y L f\thinspace w_Y \\ \vdots \end{bmatrix}}_{\mathcal{E}_Y^B} \cdot \underbrace{\begin{bmatrix} Q \end{bmatrix}}_{\mathcal{Q}}
\end{split}
    \label{eq:flux_eqn}
\end{equation}
where $f$ refers to the weight function in Eq. (\ref{eq:sp_avg}), such as $f \equiv 1$ and $f \equiv \exp[-i \zeta x_p]$ for spatial average and de-phased average, respectively; $\mathcal{A}_Y$ and $\mathcal{E}_Y^A$ denote coefficient matrices for solution vector $\mathcal{S}_Y$ and internal heat source on temperature gradient and temperature, respectively; similarly, $\mathcal{B}_Y$ and $\mathcal{E}_Y^A$ refer to coefficient matrices for solution vector $\mathcal{S}_Y$ and internal heat source on heat flux and sensible heat storage, respectively. In addition, the temperature at the concentrated heat capacity location $x_p = p_Y$ should be evaluated, i.e., $T_Y(p_Y) = \mathcal{B}_Y(p_Y) \mathcal{S}_Y + \mathcal{E}_Y^B(p_Y) \mathcal{Q}$. Since $NP$ sampling points are employed to evaluate the (weighted) spatial average in Eq. (\ref{eq:sp_avg}), it can be obtained by summing results at each sampling point multiplied by the corresponding factor, which provides quantities linearly dependent on the internal heat source only, 

\begin{equation}
\begin{aligned}
    \begin{bmatrix} \langle T_{,x} \rangle_Y \\ \langle T \rangle_Y \end{bmatrix} = \frac{1}{NP} \sum_{i=1}^{NP} \left[ (\mathcal{A}_Y)_{ip} \thinspace (\mathcal{S}_Y)_p + (\mathcal{E}_{Y}^A)_{ip} (\mathcal{Q})_p \right] = \overline{\mathcal{A}_Y} \mathcal{Q} \\ \begin{bmatrix} \langle q \rangle_Y  \\ \langle F \rangle_Y \end{bmatrix} = \frac{1}{NP} \sum_{i=1}^{NP} \left[ (\mathcal{B}_Y)_{ip} \thinspace (\mathcal{S}_Y)_{p} + (\mathcal{E}^{B}_Y)_{ip}  (\mathcal{Q})_p \right] + \begin{bmatrix}
       0 \\ H T_Y(p_Y)  
    \end{bmatrix} = \overline{\mathcal{B}_Y} \mathcal{Q} 
\end{aligned}
\label{eq:post_mat}
\end{equation}
where the solution vector $\mathcal{S}_Y$ has been replaced by $(\mathcal{F}_Y)^{-1} \mathcal{Q}$ in Eq. (\ref{eq:sol_Q}). Therefore, the realization-dependent averaged thermal quantities have been correlated to the sure internal heat source, and $\overline{\mathcal{A}_Y}$ and $\overline{\mathcal{B}_Y}$ are two corresponding coefficient matrices. Eq. (\ref{eq:post_mat}) establishes mapping between the source and averaged quantities, which allows the identification of effective properties.

\subsubsection{Effective properties of the realization variable Y}
Given realization $Y$, we will evaluate four effective properties, $\tilde{C}^\text{eff}, \tilde{\chi}^\text{eff}, \tilde{\xi}^\text{eff}$, and $\tilde{K}^{\text{eff}}$ in this subsection. In general, separating the four properties requires two independent pieces of information. For instance, it can be two linearly independent source vectors, so that the matrix $\overline{\mathcal{A}_Y}$ is square and invertible. When the $\mathcal{Q}$ has at least two independent components in Eq. (\ref{eq:post_mat}), the effective properties can be evaluated as, 

\begin{equation}
    \begin{bmatrix} \langle q \rangle_Y  \\ \langle F \rangle_Y \end{bmatrix} = \underbrace{\overline{\mathcal{B} _Y} \left(\overline{\mathcal{A}_Y}\right)^{-1}}_{\text{Effective Tensor}}     \begin{bmatrix} \langle T_{,x} \rangle_Y \\ \langle T \rangle_Y \end{bmatrix} 
    \label{eq:effec_factor}
\end{equation}
where $\overline{\mathcal{B} _Y} \left(\overline{\mathcal{A}_Y}\right)^{-1}$ is the effective constitutive tensor. In general, the effective properties in Eq. (\ref{eq:effec_factor}) are local, which depend on the realization $Y$. However, for the commonly de-phased average of weight function $f(x) = \exp [-i \zeta x]$ under the Bloch-form boundary conditions, the averaged temperature and temperature gradient are not independent. We can use the de-phased quantities to extract the effective properties \cite{Srivastava2011}. 

\subsubsection{Effective properties for de-phased averaged quantities}
Using the integral by parts with the Bloch-form boundary conditions, we obtain
\begin{equation}
    \langle T_{,x} \rangle_Y = \underbrace{\frac{1}{2L} \left(T_Y(L) \exp[-i \zeta L] - T_Y(-L) \exp[i \zeta L]\right)}_{\text{zero}} + \frac{i \zeta }{2L} \int_{-L}^{L} \exp[-i \zeta x] T_Y(x) \thinspace dx = i \zeta \langle T \rangle_Y
\end{equation}
and therefore, the rank of the matrix $\overline{\mathcal{A}_Y}$ is $1$, 
which is not invertible. Under the assumption of of the Bloch-form boundary conditions with the de-phased average of weight function $\exp[-i \zeta x]$, thermal quantities can be written as $2L$-periodic functions. For instance, define $\theta_Y(x) = T_Y(x) \exp[-i \zeta x]$, which is the $2L$-periodic function, i.e. $\theta_Y(x) = \theta_0(x - Y)$. Therefore, $T_Y(x) = \exp[i \zeta (x-Y)] \theta_0(x - Y) = \exp [-i \zeta Y] T_0(x)$. Now, the de-phased average of temperature can be expressed as, 

\begin{equation}
\begin{split}
    \langle T \rangle_Y & = \frac{1}{2L} \int_{-L}^{L} T_Y(x) \exp[-i \zeta x] \thinspace dx = \frac{1}{2L} \int_{-L}^{L} T_0(x-Y) \exp[-i \zeta (x-Y)] \thinspace dx \\ & = \frac{1}{2L} \int_{-L-Y}^{L-Y} T_0(s) \exp[-i \zeta s] \thinspace ds = \frac{1}{2L} \int_{-L-Y}^{L-Y} \theta_0(s) \thinspace ds
\end{split}
    \label{eq:dephase_shift}
\end{equation}
where $s = x - Y$. Since $\theta_0(x)$ is the $2L$-periodic function, its integral is constant when the integral interval is $2L$. Therefore, when the periodic composite is subjected to the Bloch-form boundary condition, the de-phased average remains constant for any realizations $Y$, and thus the extracted effective properties must be the same. This implies that under the Bloch-form boundary condition and de-phased average, the effective properties for any realization $Y$ are numerically equivalent to the results of the ensemble average in the following subsection. Note that the  physical meanings are different: a single realization $Y$ generally provides local effective properties, while the ensemble average yields non-local effective descriptions, which will be elaborated subsequently. 

To separate the four effective properties from the de-phased averaged thermal quantities, this subsection considers two cases at the same frequency but with opposite macroscopic thermal wavenumbers, such as $\zeta$ and $-\zeta$. The correctness of this scheme will be justified later in Eq. (\ref{eq:effective_properties_fourier}) from the explicit Fourier-series formulae. With the extraction scheme, one obtains, 

\begin{equation}
\begin{aligned}
    \langle q \rangle_Y = -K_Y^{\text{eff}} (i \zeta) \langle T \rangle_Y + \chi_Y^{\text{eff}} \langle T \rangle_Y \\ 
    \langle F \rangle_Y = \xi_Y^{\text{eff}} (i \zeta) \langle T \rangle_Y + C_Y^{ \text{eff}} \langle T \rangle_Y \\
\end{aligned}
    \label{eq:effec_bloch_Y}
\end{equation}
and 
\begin{equation}
\begin{aligned}
    m_1^Y = \frac{\langle q \rangle_Y(\zeta)}{\langle T \rangle_Y(\zeta)} = -K_Y^\text{eff} (i\zeta) + \chi_Y^\text{eff}, \quad m_2^Y = \frac{\langle q \rangle_Y(-\zeta)}{\langle T \rangle_Y(-\zeta)} = K_Y^\text{eff} (i\zeta) + \chi_Y^\text{eff} \\
    n_1^Y = \frac{\langle F \rangle_Y(\zeta)}{\langle T \rangle_Y(\zeta)} = C_Y^\text{eff} + \xi^\text{eff} (i\zeta), \quad n_2^Y = \frac{\langle F \rangle_Y(-\zeta)}{\langle T \rangle_Y(-\zeta)} = C_Y^\text{eff} - \xi_Y^\text{eff}(i\zeta) 
    \end{aligned}
\end{equation}
Hence, the effective properties are, 

\begin{equation}
    K_Y^{\text{eff}} = -\frac{m_1^Y - m_2^Y}{2 i \zeta}, \quad \chi_Y^{\text{eff}} = \frac{m_1^Y + m_2^Y}{2}, \quad \xi_Y^{\text{eff}} = \frac{n_1^Y - n_2^Y}{2 i \zeta}, \quad C_Y^{\text{eff}} = \frac{n_1^Y + n_2^Y}{2}
    \label{eq:effec_Y}
\end{equation}

\subsubsection{Effective properties of the ensemble average}
Following Eqs. (\ref{eq:en_avg}) and  (\ref{eq:post_mat}), the ensemble average with respect to the realization variable $Y$ can be interpreted as the expectation operation. To avoid the ambiguity between the spatial (de-phased) average and the ensemble average, it is emphasized that $\langle \cdot \rangle_Y$ refers to the de-phased average over the unit cell with the realization $Y$, while $\langle \cdot \rangle$ stands for the ensemble-averaged quantity. For instance, when the probability function $P(Y) = \frac{1}{2L}$ is uniform, the expectation of the de-phased average of temperature becomes, 

\begin{equation}
\begin{split}
\mathbb{E} \left[ \langle T \rangle_Y \right]
&=\frac{1}{2L} \int_{-L}^{L}\int_{-L}^{L} T_Y(x)\,e^{-i\zeta x}\,dx\,dY =\frac{1}{2L}\int_{-L}^{L} \langle T\rangle(x)\,e^{-i\zeta x}\,dx
\end{split}
\label{eq:temp_en_spatial}
\end{equation}
where the interchange of integration sequence shows that the expectation of the de-phased average is equivalent to the de-phased average of the ensemble-averaged temperature. In the following, we clarify why the ensemble-averaged operators can be regarded as convolution kernels and subsequently evaluated using the Fourier series. Under the random realization $Y$ and Bloch-form boundary conditions, the microstructure-specific Green's function satisfies the shift relation $G_Y(x, x') = G_0(x-Y, x'- Y)$. When the Green's function is averaged over uniformly distributed realizations, the ensemble-averaged kernel only depends on the relative distance $x - x'$, which implies that the ensemble-averaged Green's function and its related kernels are translation-invariant convolution kernels. Thanks to such a property, the spatial convolution operation in the effective constitutive relations can be transformed into multiplication in the Fourier space. For Bloch-form thermal fields, the de-phased average removes the macroscopic phase factor ($\zeta$) and extracts the averaged quantity corresponding to the same Bloch component. With a uniform probability density function, the ensemble average can be numerically approximated with $NY$ realizations, i.e., $Y_i \in [-L, L]$ are $NY$ uniform sampling points, 

\begin{equation}
\begin{aligned}
    \begin{bmatrix} \langle \widetilde{T_{,x}} \rangle \\ \langle \tilde{T} \rangle \end{bmatrix} = \frac{1}{NY} \sum_{i=1}^{NY} \begin{bmatrix} \langle T_{,x} \rangle_i \\ \langle T \rangle_i \end{bmatrix} = \frac{1}{NY} \sum_{i=1}^{NY} \overline{\mathcal{A}}^i \mathcal{Q} = \langle \overline{\mathcal{A}} \rangle \mathcal{Q}
    \\%
     \begin{bmatrix} \langle \widetilde{q} \rangle \\ \langle \tilde{F} \rangle \end{bmatrix} = \frac{1}{NY} \sum_{i=1}^{NY} \begin{bmatrix} \langle q \rangle_i  \\ \langle F \rangle_i \end{bmatrix} = \frac{1}{NY} \sum_{i=1}^{NY}\overline{\mathcal{B}}^i \mathcal{Q} = \langle \overline{\mathcal{B}} \rangle \mathcal{Q}
\end{aligned}
    \label{eq:ensem_mat}
\end{equation}
where $\langle \overline{\mathcal{A}} \rangle$ and $\langle \overline{\mathcal{B}} \rangle$ are two ensemble-averaged coefficient matrices. Since the internal heat source is sure, the ensemble average operation is only applied on the coefficient matrix, which agrees with Eq. (\ref{eq:thermal_ens}). In numerical evaluation, $Y_i$ are sampled uniformly in $[-L, L]$, and a convergence test on $NY$ is provided in Section 4.4 of the Supplemental Material.

Although the ensemble average of temperature and temperature gradient are not independent under the Bloch-form and de-phased average settings, Eqs. (\ref{eq:effec_bloch_Y})-(\ref{eq:effec_Y}) showed that they can be obtained through the odd and even properties. Moreover, Eq. (\ref{eq:dephase_shift}) shows the shift-invariance, which implies that the effective properties for the ensemble average and a single realization should be the same. Hence, the similar procedure is not repeated below. For other general weight functions, one can follow the same procedure in Eq. (\ref{eq:effec_factor}) as well.

\section{The Exact homogenization method}
Although recent works have proposed the bi-anisotropic effective constitutive relations for heat transfer, the present formulae  provides the first explicit demonstration of nonlocal thermal Willis kernels for periodic laminates. Such results serve as a benchmark for validating the homogenization scheme in Section 3 and the boundary retrieval method. 

\subsection{Method of Green's function}
Given the specific microstructure and boundary conditions, the local solution of Eq. (\ref{eq:heat_eqn_harm}) can be expressed formally by the Green's function as follows \cite{Willis2009}: 

\begin{equation}
    T_Y(x) = G_Y * Q_Y = \int_{-L}^{L} G_Y(x, x') \thinspace Q_Y(x') \thinspace dx' \quad \text{and} \quad T_{Y,x} = G_{Y,x} * Q_Y = \int_{-L}^{L} G_{Y,x}(x, x') \thinspace Q_Y(x') \thinspace dx'
\end{equation}
where $G_Y(x, x')$ is a specific expression of Green's function defined for the microstructure subjected to the boundary conditions with realization $Y$. When the microstructure is periodic, Green's function satisfies the condition that $G_{Y}(x, x') = G_{0}(x - Y, x' - Y)$, where $G_{0}$ refers to the Green's function with the realization of $Y = 0$. The source field can be prescribed by distribution $Q(x)$ with a weight function $w_Y^{(1)}(x)$, i.e., $Q_Y(x) = w_{Y}^{(1)}(x) Q(x)$. Hence, the source field becomes $Y$-dependent even though the initial heat source field is invariant. Moreover, Willis \cite{Willis2009} commented that the measurement of average fields may become unavailable for voids, and an auxiliary weight function $w_{Y}^{(2)}(x)$ can be introduced to handle it. For instance, $w_{Y}^{(2)} \equiv 1$ for general cases and $w_{Y}^{(2)} = 1 / c_2$ when the measurement of material phase 1 is void. For the illustration purposes, this paper consider $w_Y = w_Y^{(1)} = w_Y^{(2)}$. Using the local constitutive relations, the heat flux and sensible heat change can be expressed as, 

\begin{equation}
    q_Y(x) = -K_Y(x)\int_{-L}^{L} G_{Y,x}(x, x') \thinspace Q_Y(x') dx' \quad \text{and} \quad F_Y(x) = C_Y (x)\int_{-L}^{L} G_Y(x, x') \thinspace Q_Y(x') dx'
\end{equation}
Since the ensemble average is of interest, some properties can be utilized to simplify the expressions. For instance, the ensemble average of weighted temperature is, 

\begin{equation}
\begin{split}
    \langle w T \rangle (x) & = \langle w \thinspace G * Q \rangle (x) = \frac{1}{2L} \int_{-L}^L \int_{-L}^{L} w_Y(x) G_Y(x, x') w_Y(x') Q(x') \thinspace dx' \thinspace dY \\
    & =  \int_{-L}^L \left[ \frac{1}{2L} \int_{-L}^{L} w_Y(x) G_Y(x, x') w_Y(x') \thinspace dY \right] \thinspace Q(x') \thinspace dx' = \langle w G w \rangle * Q
\end{split}
    \label{eq:ensemble_quantity}
\end{equation}
Following the same procedure, other ensemble-averaged thermal quantities can be obtained as, 

\begin{equation}
    \langle wT_{,x} \rangle = \langle w G_{,x} w\rangle * Q \quad \text {and} \quad \langle q \rangle = -\langle K G_{,x} w\rangle * Q \quad \text{and} \quad \langle F \rangle  = \langle C G w\rangle * Q
    \label{eq:thermal_ens}
\end{equation}
Here, the sure heat source $Q = \langle w G w \rangle^{-1} * \langle w T \rangle$ can be applied, which is the deconvolution process. Note that the heat source, whether sure or weighted, is a transitional quantity that will not be evaluated during the process. Therefore, Green's function and the source-driven method are independent of prescribed sources. 

\subsection{Explicit Green's function of the microstructure}
This subsection derives the explicit Green's function for the microstructure presented in Section 2, assuming periodic thermal properties, which is analogous to the elastodynamic case \cite{Willis2009}. The method can be extended to other periodic microstructures. Without loss of generality, we derive Green's function with the realization $Y = 0$, $G_{Y=0}(x, x')$ for the governing equation:
\begin{equation}
    \left[ K_{Y=0}(x) G_{Y=0,x}(x,x') \right]_{,x} + i \omega C_{Y=0} G_{Y=0}(x, x') = -\delta(x - x')
    \label{eq:GreenPDE}
\end{equation}
Following the elastodynamic case \cite{Willis2009}, we write the solution in the Floquet form, 

\begin{equation}
    \phi_+ (x) = \exp[\mu x] \psi_+(x), \quad \phi_- (x) = \exp[-\mu x] \psi_-(x)
\end{equation}
where $\psi_\pm(\textbf{x})$ are periodic functions over period $2L$; $\mu$ is the Floquet number of the microstructure-specific Green's function, which is subsequently determined by the frequency and thermal properties. Note that $\mu$ characterizes the attenuation and variation of the Green's function and is different from thermal wavenumber $\zeta$ in the Bloch wave of macroscopic harmonic heat transfer. Due to the decaying feature of the heat equation, the variable $\mu$ is assumed to have a positive real part. The Green's function $G_{Y=0}(x, x')$ becomes,

\begin{equation}
    G_{Y=0}(\textbf{x}, \textbf{x}') = \begin{cases} D \phi_+(x) \phi_-(x') & x < x' \\ D \phi_+(x') \phi_-(x) & x > x'\end{cases}
    \label{eq:Gfunc}
\end{equation}
Since the Green's function should satisfy the continuity condition except that an inflection point exists at $x = x'$  due to the Dirac delta function in Eq. \eqref{eq:GreenPDE}, the coefficient $D$ can be obtained as, 

\begin{equation}
    D = \frac{1}{K_{Y=0}(x)} \left[ \phi'_+(x) \phi_-(x) - \phi_+(x) \phi'_-(x) \right]^{-1}
\end{equation}
Note that although the above equation shows that $D$ is a function of $x$, the results in Section 5 show that $D$ is independent of $x$. In the following, the component $\phi_+(x)$ is derived. Because the other component $\phi_-(x)$ shares a similar process, the derivation process will not be repeated below. The component function $\phi_+$ can be constructed in terms of four branches considering continuity and periodicity, 

\begin{equation}
    \phi_+(x) = \begin{cases}
    \exp[-\mu L] \left( A \cosh[k_2 (L + x)] + B \sinh [k_2 (L + x)] \right) & x \in [-L, -c_1 L) \\ 
    \cosh \left[ k_1 x \right] + b \sinh \left[ k_1 x \right]  & x \in [-c_1 L, \alpha c_1 L) \\
     c \cosh \left[ k_1 x \right] + d \sinh \left[ k_1 x \right] & x \in [\alpha c_1 L, c_1 L)\\
    \exp[\mu L] \left( A \cosh \left[ k_2 (L - x) \right] - B \sinh \left[ k_2 (L - x) \right] \right) & x \in [c_1 L, L]
    \end{cases}
    \label{eq:form_sol}
\end{equation}
where $k_1 = \frac{\sqrt{2}}{2} (1 - i) \sqrt{\omega \frac{C^1}{K^1}}$ and $k_2 = \frac{\sqrt{2}}{2} (1 - i) \sqrt{\omega \frac{C^2}{K^2}}$, and coefficients $A, B, b, c, d$ and $\mu$ are determined by continuity conditions and the jump condition at the concentrated heat capacity, and the determination of coefficients and the Floquet number $\mu$ for the specific microstructure are elaborated in the Supplemental Material 1.1. The other branch $\phi_-$ can be obtained in the same manner by using the corresponding Floquet number $-\mu$, while all equations on coefficients are retained. 

\subsection{Ensemble averages of Green's function}
The following derives the ensemble averages of thermal quantities. Although the explicit Green's function can be obtained in Eq. (\ref{eq:Gfunc}), the direct application of the piecewise function in Eq. (\ref{eq:form_sol}) to the ensemble average over different realizations $Y$ is cumbersome. The primary reason is that the realization $Y$ shifts the microstructure, and it subsequently changes the microstructure-specific Green's function, which causes repeated evaluation of $\phi^+$ and $\phi_-$. Therefore, the explicit Green's function is written in a Fourier-series form, which mathematically accounts for the periodicity and shift dependence of the Green's function. In Eqs. (\ref{eq:en_avg}) and (\ref{eq:ensemble_quantity}), the ensemble average can be computed from Green's function and the associated weight functions. In Eq. (\ref{eq:Gfunc}), the Green's function is expressed in terms of the exponential function and two periodic functions $\psi_\pm$. Hence, it is rational to express Green's function in terms of the Fourier series: 

\begin{equation}
    G_Y(x, x') = D \begin{cases} \exp[\mu(x - x')] \sum_{m=-\infty}^{\infty} a_m(\mu) \exp\left[ \frac{-i m \pi (x- Y)}{L} \right] \sum_{n = -\infty}^{\infty} a_n(-\mu) \exp \left[ \frac{-i n \pi (x' - Y)}{L} \right] & x < x' \\
    \exp[-\mu(x - x')] \sum_{m=-\infty}^{\infty} a_m(\mu) \exp\left[ \frac{-i m \pi (x'- Y)}{L} \right] \sum_{n = -\infty}^{\infty} a_n(-\mu) \exp \left[ \frac{-i n \pi (x - Y)}{L} \right] & x > x'
    \end{cases}
    \label{eq:Gfunc_series}
\end{equation}
where Fourier coefficients $a_m(\mu)$ and $a_{m}(-\mu)$ are associated with the periodic part of Green's function, and the detail derivation is seen in Appendix A. Subsequently, the ensemble average of Green's function can be derived using the orthogonal properties of trigonometric functions, 

\begin{equation}
    \langle G \rangle = \frac{1}{2 L} \int_{-L}^{L} G_Y(x, x') \thinspace dY = D \exp [-\mu|x - x'|] \sum_{m=0}^{\infty} a_m(\mu) a_{-m}(-\mu) \exp \left[ \frac{i m \pi |x - x'|}{L} \right]
    \label{eq:Gfunc_ens}
\end{equation}

In addition to the ensemble average of Green's function, the ensemble average of weighted Green's function is obtained as, 

\begin{equation}
    \langle w G w \rangle = D \exp[-\mu |x - x'|] \sum_{m = -\infty}^{\infty} a_m^w(\mu) a_{-m}^w(-\mu) \exp \left[ \frac{i m \pi |x - x'|}{L} \right]
    \label{eq:Gfunc_weight_ens}
\end{equation}
where two additional Fourier coefficients $a_m^w(\mu)$ and $a_m^w(-\mu)$ are provided in Appendix A as well. Although the Fourier series form of Green's function and its ensemble average are expanded to infinite terms, the convergence tests in Supplemental Material Section 4.3 show that 50 terms can provide highly accurate results. This paper uses 100 terms. 

\subsection{Ensemble averages of thermal quantities}
Section 3. (a) shows that the primary concept of the source-driven method is to link all thermal quantities with the source field, such as $Q = \langle w G w \rangle^{-1} * \langle w T \rangle$. Hence, the evaluation of ensemble-averaged heat flux and sensible heat change requires $\langle K G_{,x} w \rangle$ and $\langle C G w \rangle$. Since the local microscopic thermal properties and weight functions are periodic, they can be expressed in terms of the Fourier series as follows: 

\begin{equation}
\begin{aligned}
    &-K_Y(x) G_{Y,x}(x,x') w_Y(x') = \\
    &D \exp[-\mu|x - x'|] \begin{cases} \sum_{m = -\infty}^{\infty} b_{m}(\mu) \exp \left[ \frac{-im\pi(x-Y)}{L} \right]  \sum_{n = -\infty}^{\infty} a_n^w(-\mu) \exp \left[ \frac{-in\pi(x'-Y)}{L} \right] & x < x' \\
    \sum_{m = -\infty}^{\infty} b_{m}(-\mu) \exp \left[ \frac{-im\pi(x-Y)}{L} \right] \sum_{n = -\infty}^{\infty} a_m^w(\mu) \exp \left[ \frac{-in\pi(x'-Y)}{L} \right]
    & x > x'
    \end{cases}
\end{aligned}
\end{equation}
and $C G w$ can be expressed in a similar manner, 

\begin{equation}
\begin{aligned}
    &C_Y(x) G_Y(x, x') w_Y(x') =\\
    &D \exp[-\mu|x - x'|] \begin{cases} \sum_{m = -\infty}^{\infty} c_{m}(\mu) \exp \left[ \frac{-im\pi(x-Y)}{L} \right]  \sum_{n = -\infty}^{\infty} a_n^w(-\mu) \exp \left[ \frac{-in\pi(x'-Y)}{L} \right] & x < x' \\
    \sum_{m = -\infty}^{\infty} c_{m}(-\mu) \exp \left[ \frac{-im\pi(x-Y)}{L} \right] \sum_{n = -\infty}^{\infty} a_n^w(\mu) \exp \left[ \frac{-in\pi(x'-Y)}{L} \right] & x > x'
    \end{cases}
\end{aligned}
\end{equation}
where four Fourier coefficients $b_m(\mu), b_m(-\mu), c_m(\mu)$ and $c_m(-\mu)$ are provided in Appendix A. (ii). The ensemble average of $-\langle K G_{,x} w \rangle$ and $-i\omega \langle C G w \rangle$ can be obtained as, 

\begin{equation}
    -\langle K G_{,x} w \rangle = D \exp[-\mu |x - x'|] \begin{cases}
        \sum_{m=-\infty}^{\infty} b_m(\mu) a_{-m}^w(-\mu) \exp \left[ \frac{i m \pi |x - x'|}{L} \right] & x < x' \\
        \sum_{m=-\infty}^{\infty} a_m^w(\mu) b_{-m}(-\mu) \exp \left[ \frac{i m \pi |x - x'|}{L} \right] & x > x'
    \end{cases}
    \label{eq:flux_ens}
\end{equation}
and 

\begin{equation}
    -i \omega \langle C G w \rangle = D \exp[-\mu |x - x'|] \begin{cases}
        \sum_{m=-\infty}^{\infty} c_m(\mu) a_{-m}^w(-\mu) \exp \left[ \frac{i m \pi |x - x'|}{L} \right] & x < x' \\
        \sum_{m=-\infty}^{\infty} a_m^w(\mu) c_{-m}(-\mu) \exp \left[ \frac{i m \pi |x - x'|}{L} \right] & x > x'
    \end{cases}
    \label{eq:heat_ens}
\end{equation}

\subsection{Extraction of effective constitutive relations}
The subsection derives the effective constitutive relations based on the ensemble-averaged thermal quantities. As previously discussed in Section 3. (a), the primary concept of the Green's function and the source-driven method is to bridge thermal quantities with the heat sources. Substituting $Q = \langle w G w \rangle^{-1} * \langle w T \rangle$ into Eq. (\ref{eq:thermal_ens}), the ensemble-averaged heat flux and sensible heat change can be obtained using the weighted temperature, 

\begin{equation}
    \langle q \rangle = -\langle K G_{,x} w \rangle * \langle w G w \rangle^{-1} * \langle w T \rangle, \quad \langle F \rangle = \langle C G w \rangle * \langle w G w \rangle^{-1} * \langle w T \rangle
    \label{eq:conv_thermal}
\end{equation}
Note that the ensemble average of operators, (weighted) Green's function, only depends on distance $x - x'$. Therefore, the convolution operations can be transformed into multiplication in the Fourier space, 

\begin{equation}
    \langle \tilde{q} \rangle(\kappa) = -\langle \widetilde{K G_{,x} w} \rangle \langle \widetilde{w G w} \rangle^{-1} \langle \widetilde{wT} \rangle, \quad \langle \tilde{F} \rangle(\kappa) = \langle \widetilde{C G w} \rangle \langle \widetilde{w G w} \rangle^{-1} \langle \widetilde{wT} \rangle
    \label{eq:conv_thermal_fourier}
\end{equation}
where  $\widetilde{(.)}$ refers to the Fourier transform in terms of the variable $\kappa$ in the Fourier space, and the latter is used to improve readability for longer terms. The variable $\kappa$ refers to the macroscopic wavenumber. Specifically, when the unit cell (length $2L$) is subjected to the Bloch-form boundary condition, the admissible macroscopic wavenumber becomes $\kappa = \zeta + \frac{n \pi}{L}$, where $\zeta \in [-\frac{\pi}{L}, \frac{\pi}{L})$ and $\frac{n \pi}{L}$ ($n \in \mathcal{Z}$) is the one-dimensional reciprocal lattice number. 

With some straightforward derivation in Supplemental Material 1.3, the effective thermal properties in the Fourier space can be extracted as:

\begin{equation}
\begin{aligned}
    \tilde{K}^{\text{eff}} = -D \sum_{m = -\infty}^{\infty} \left[ \frac{b_m(\mu) a_{-m}^w(-\mu) - a_m^w(\mu) b_{-m}(-\mu)}{\beta_m^2 + \kappa^2} \right] \langle \widetilde{w G w} \rangle^{-1} \\
    \tilde{\chi}^{\text{eff}} = D \sum_{m = -\infty}^{\infty} \left[\beta_m \frac{a_m^w(\mu) b_{-m}(-\mu) + b_m(\mu) a_{-m}^w(-\mu)}{\beta_m^2 + \kappa^2} \right] \langle \widetilde{w G w} \rangle^{-1} \\
    \tilde{\xi}^{\text{eff}} = D \sum_{m = -\infty}^{\infty} \left[ \frac{c_m(\mu) a_{-m}^w(-\mu) - a_{m}^w(\mu) c_{-m}(-\mu)}{\beta_m^2 + \kappa^2} \right] \langle \widetilde{w G w} \rangle^{-1}\\
    \tilde{C}^{\text{eff}} = D \sum_{m = -\infty}^{\infty} \left[ \beta_m \frac{a_m^w(\mu) c_{-m}(-\mu) + c_{m}(\mu) a_{-m}^w(-\mu)}{\beta^2_m + \kappa^2} \right] \langle \widetilde{w G w} \rangle^{-1}
\end{aligned}
\label{eq:effective_properties_fourier}
\end{equation}
Note that using Eqs. (\ref{eq:c_coef_2}) and  (\ref{eq:c_coef_2m}), we can correlate coupling terms $\tilde{\chi}^{\text{eff}}$ and $\tilde{\xi}^{\text{eff}}$ as follows: 

\begin{equation}
    \tilde{\xi}^{\text{eff}} = D \sum_{m = -\infty}^{\infty} \left[ \frac{1}{i \omega} \beta^m  \frac{a_m^w(\mu) b_{-m}(-\mu) + b_m(\mu) a_{-m}^w(-\mu)}{\beta_m^2 + \kappa^2}\right] = \frac{1}{i \omega} \tilde{\chi}^{\text{eff}} 
\end{equation}

\section{Numerical case studies}
To demonstrate the formulation, we conduct numerical case studies with the following  material properties: (i) $C^1 = 0.5 J / m^3 \cdotp K, K^1 = 0.05 W / m \cdotp K$; (ii) $C^2 = 0.5 J / m^3 \cdotp K, K^2 = 1 W / m \cdotp K$; (iii) the length $2L = 2$ m, $\alpha = 0.75$, and the concentrated heat capacity $H = 2 L C^1 = 1 J / m^2 \cdotp K$. The frequency is normalized as $\overline{\omega} = (2L)^2 \omega / (K^2 / C^2) = 2 \omega$. For ensemble averages, $NY = 100$ uniformly distributed sampling points are adopted. 

\subsection{Green's function}
\begin{figure}
    \centering
    \includegraphics[width=\linewidth]{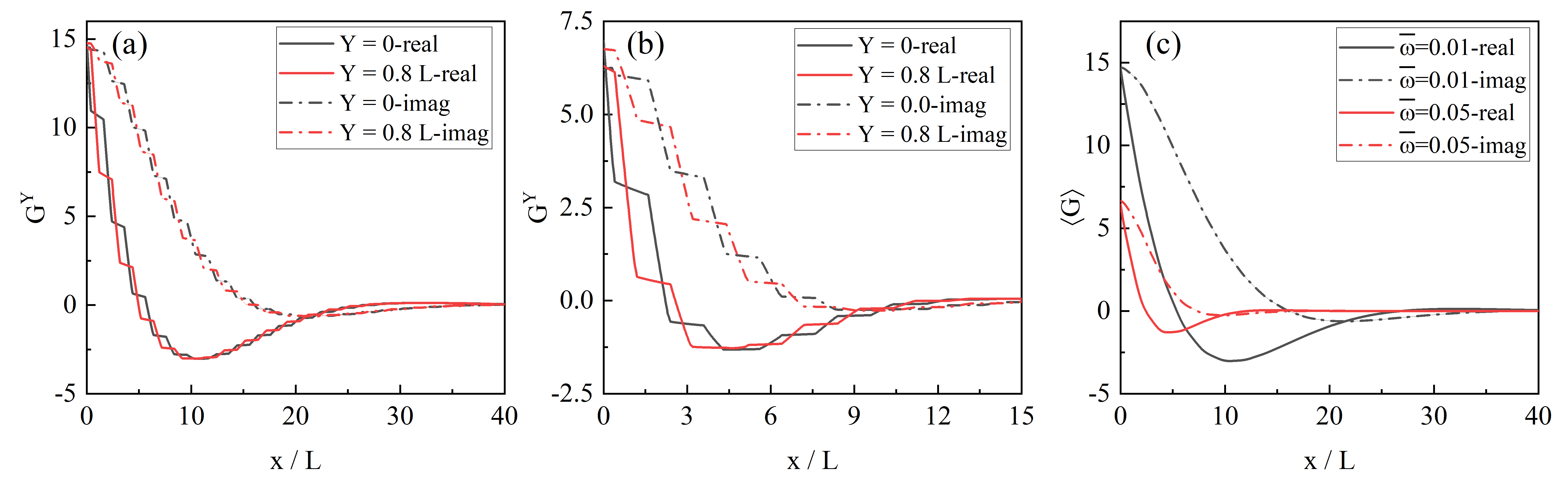}
    \caption{Variation of Green's function with normalized frequency $\overline{\omega} = 0.01$ or $0.05$, (a) Green's function with realization $Y = 0, 0.8 L$, $\overline{\omega} = 0.01$ when $x / L \in [0, 40]$; (b) Green's function with realization $Y = 0, 0.8 L$, $\overline{\omega} = 0.05$ when $x / L \in [0, 15]$; and (c) unweighted ensemble-averaged Green's function.}
    \label{fig:Gfunc}
\end{figure}

Figs. \ref{fig:Gfunc} (a) and (b) plot the variation of Green's function of the specific microstructure, when the normalized frequency $\overline{\omega} = 0.01, 0.05$, respectively. As Eq. (\ref{eq:Gfunc}) indicates, the Green's function is composed of a coefficient $D$ and two component functions. Specifically, when the normalized frequency is given as $\overline{\omega} = 0.01$ or $0.05$, the coefficient $D$ is $14.841 + 14.488 i, 6.973 + 6.186 i$, and the Floquet number $-0.147 + 0.146 i$ or $-0.332 + 0.323 i$, respectively. In general, the coefficient $D$ and the Floquet number affect the magnitude and damping distance of the Green's function at given frequencies, which are intrinsic properties of the specific microstructure. 

Figs. \ref{fig:Gfunc} (a) and (b) show that the Green's function is dependent on the realization of the microstructure. The cases $Y = 0$ and $Y = 0.8 L$ exhibit apparent discrepancies in the first few unit cells, i.e., $x / L < 10$, with the primary differences occurring in the peak values and local oscillations of both the real and imaginary parts. However, when $x / L$ increases, the differences between the two realizations reduce. This phenomenon indicates that the realization parameter $Y$ significantly influences short-range responses, whereas the Floquet number governing long-range responses remains unchanged. This explains that the long-range response is governed by the microstructure's intrinsic features rather than by the unit-cell endpoints. Another noticeable feature in Figs. \ref{fig:Gfunc} (a) and (b) is the piecewise variations of Green's function, which mainly arise from the heterogeneous configuration of the microstructure. Since the material properties are piecewise constant in Eq. (\ref{eq:mat_phase}), the slope variation of Green's function indicates the change of thermal conductivity. Despite differences in local responses, Green's function exhibits an exponentially damped trend, governed by the Floquet number. As previously mentioned, when $\overline{w} = 0.05$, the real part of the Floquet number is greater, which is consistent with the faster damping trend in Fig. \ref{fig:Gfunc} (b).  

Fig. \ref{fig:Gfunc} (c) presents the ensemble average of Green's function $\langle G \rangle$, in which the local responses are smoothed through averaging over realizations. Compared to two specific realizations in Fig. \ref{fig:Gfunc} (a) and (b), two observations are: (i) the magnitude of Green's function is close; and (ii) the ensemble-averaged Green's function becomes much smoother, while the exponentially decaying feature is retained. It can be concluded that although the ensemble-averaged Green's function is independent of realizations, it is still governed by the coefficient $D$ and the Floquet number, which retains the intrinsic features of the microstructure. 

\begin{figure}
    \centering
    \includegraphics[width=0.7 \linewidth]{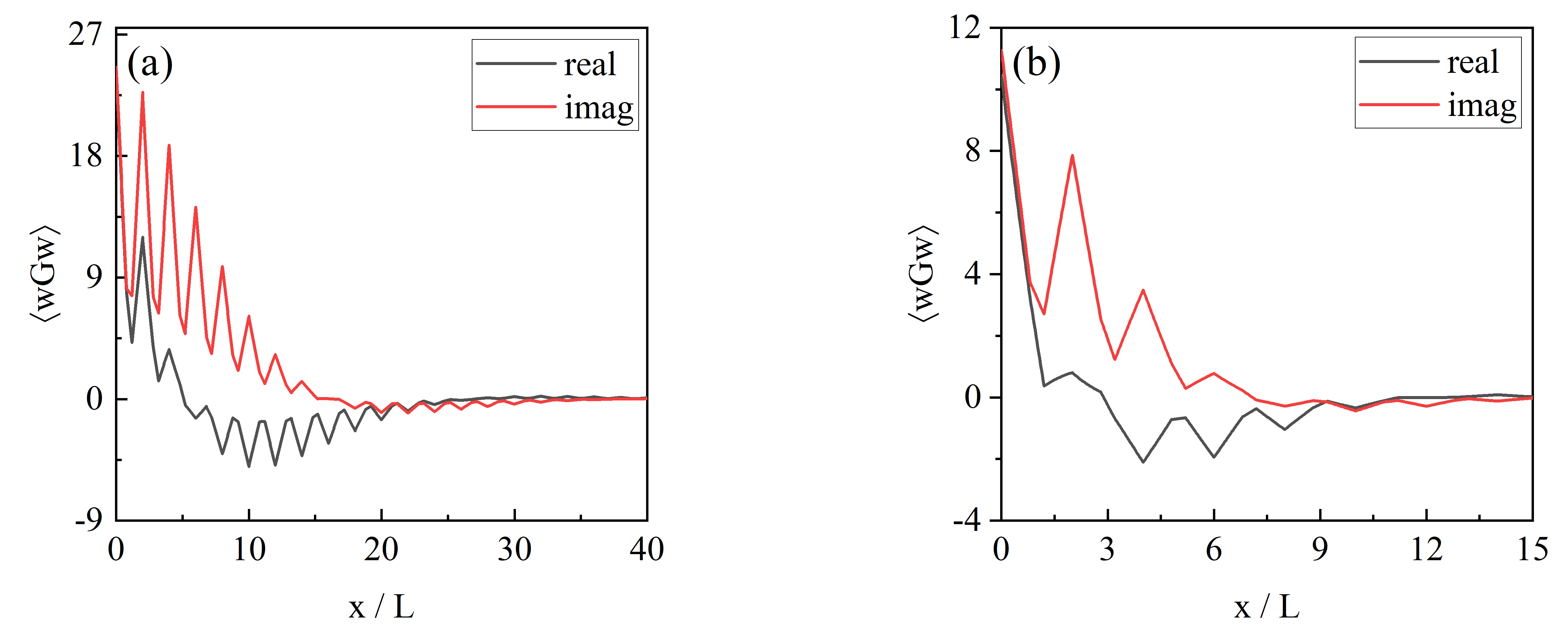}
    \caption{Variation of weighted ensemble-averaged Green's function $\langle w G w \rangle$, (a) $\overline{\omega} = 0.01$ and $x / L \in [0, 40]$; and (b) $\overline{\omega} = 0.05$ and $x / L \in [0, 15]$. The weight function $w = 1$ in phase 2 and $0$ in phase 1. }
    \label{fig:Gfunc_weighted}
\end{figure}

Fig. \ref{fig:Gfunc_weighted} (a) and (b) plots the weighted ensemble-averaged Green's function $\langle w G w \rangle$, in which the weight function is 0 in the first phase, and $1 / c_2$ in the second phase. Hence, compared to the full Green's function in Fig. \ref{fig:Gfunc} (c), the weighted ensemble-averaged Green's function should be interpreted as the ``filtering'' kernel, as the response of the first phase is suppressed. Another apparent feature is the pronounced short-range oscillation and peak-valley pattern, which is primarily caused by the discontinuous piecewise weight function. When $x$ varies, only the segment in the second phase is sampled. The sharp fluctuations in both real and imaginary parts indicate that the weighted ensemble-averaged Green's function is more sensitive to microstructural details than the unweighted case. As $x / L$ increases, the real and imaginary parts gradually decay to zero, which reveals that the Floquet number still governs the long-range response. Compared Fig. \ref{fig:Gfunc_weighted} (a) to Fig. \ref{fig:Gfunc_weighted} (b), when the normalized frequency $\overline{\omega}$ is greater, it decays faster, which is consistent with the Floquet number previously mentioned. 

Note that different weight functions enable different filtering features of ensemble-averaged quantities, including the Green's function, temperature, and temperature gradients. In other words, weight functions are tools for determining how thermal responses are observed and compared. Therefore, different weight functions generally lead to different effective constitutive relations accordingly. We will revisit this issue subsequently. 


\subsection{Comparison of homogenization schemes on unweighted average}
The following compares three homogenization schemes: (i) the EIM-powered homogenization scheme in Section 3; (ii) the exact homogenization scheme in Section 4; and (iii) the boundary-retrieval method proposed by Shmuel and Willis in \cite{Gal2025a}. Note that the boundary-retrieval method requires two groups of independent boundary responses, such as $T^L, q^L$ and $T^R, q^R$ to back calculate the effective material constants; the transfer matrix is introduced to extract effective properties, see Eq. (6.25) in \cite{Gal2025a}. Particularly, the boundary/loading conditions must be consistent when comparing homogenization schemes; therefore, the boundary responses should only be solved under the Bloch-form boundary conditions with the same internal heat sources. Following the same procedure of ensemble average, the ensemble-averaged boundary responses can be expressed as, 

\begin{equation}
    \langle T^{L,R} \rangle = \frac{1}{2L} \int_{-L}^L T_Y^{L,R}(Y) dY = \frac{1}{NY} \sum_{i=1}^{NY} T_Y^{L,R}, \quad \langle q^{L,R} \rangle = \frac{1}{2L} \int_{-L}^L q_Y^{L,R}(Y) dY = \frac{1}{NY} \sum_{i=1}^{NY} q_Y^{L,R}
    \label{eq:boundary_ens}
\end{equation}

Hence, the boundary-retrieval method still requires the local solution of thermal fields, which can be obtained by solving the global system in Eq. (\ref{eq:global_form}). To generate two independent boundary responses, it is reasonable to use $\zeta$ and $-\zeta$ as the macroscopic wavenumbers. With a non-zero internal heat source, the boundary-retrieval method identified effective properties should satisfy the ensemble-averaged heat equation, which introduces a constraint absent in the source-free boundary-retrieval method \cite{Gal2025a}, 

\begin{equation}
\begin{aligned}
    & \langle -q_{,x} \rangle + i \omega \langle F \rangle + Q = 0, \quad i\zeta \left( \tilde{K}^\text{eff} \langle T_{,x} \rangle - \tilde{\chi}^{\text{eff}} \langle T \rangle\right) + i\omega \left( \tilde{C}^\text{eff}\langle T \rangle + \tilde{\xi}^{\text{eff}}  \langle T_{,x} \rangle \right) + Q = 0
\end{aligned}
    \label{eq:ens_heat_eqn}
\end{equation}
where for deterministic internal heat source, $\langle Q \rangle = Q$; $\langle T_{,x} \rangle = i \zeta \langle T \rangle $. Hence, the effective properties $\tilde{K}^{\text{eff}}, \tilde{\xi}^{\text{eff}}, \tilde{\chi}^{\text{eff}}$ are retained (the same as Eq. (6.25) in \cite{Gal2025a}), but the $\tilde{C}^\text{eff}$ should be modified to satisfy Eq. (\ref{eq:ens_heat_eqn}) as, 

\begin{equation}
    \tilde{C}^{\text{eff}} = \frac{1}{i \omega} \left( -\frac{Q}{\langle T \rangle} + \zeta^2 \tilde{K}^\text{eff}  \right)
    \label{eq:modified_C}
\end{equation}

In the presence of an internal heat source, the boundary-retrieval method is no longer boundary-based, because the energy balance requires the knowledge of the averaged temperature. To ensure the correctness of boundary responses and ensemble-averaged temperature, the modified BRM is powered by the same EIM framework in Section 3. It should be emphasized that the effective properties (except $\tilde{C}^{\text{eff}}$) are independent of the internal states (local solutions). Hence, it cannot be applied to the weighted ensemble-averaged cases investigated in the following subsection, because weighted averages depend on internal field distributions rather than boundary responses alone. Note that the two following case studies only consider the nonzero concentrated heat capacity, and case $H = 0$ is provided in Section 5 of the Supplemental Material.

Figs. \ref{fig:zeta_0_unweight} and \ref{fig:zeta_1_unweight} plot the frequency-dependent effective properties extracted by three homogenization schemes, such as the modified boundary-retrieval (BR), the exact homogenization solution (exact), and the EIM implementation (EIM) using weight function $f(x) = \exp[-i \zeta x]$. In Figs. \ref{fig:zeta_0_unweight} and \ref{fig:zeta_1_unweight}, the curves by three homogenization schemes agree well with each other for both real and imaginary parts. The excellent agreement among three homogenization schemes demonstrates that the two proposed schemes, exact and EIM, are robust not only for the long-wavelength case ($\zeta  \to 0$) but also for a finite macroscopic thermal , which is demonstrated by the case of $\zeta  = 1$, subsequently. Particularly, when $\zeta  = 0$ , it reduces to the static limit (homogenization limit) of the layered composite. Specifically, the analytical solution of effective thermal conductivity is $0.116 W / m \cdotp K$, and the effective heat capacity is $1 J / m^3 \cdotp K$; whereas Figs. \ref{fig:zeta_0_unweight} (a) and (b) indicate that when $\overline{\omega} = 0$, $\tilde{C}^{\text{eff}} = 1 J / m^3 \cdotp K$, $\tilde{K}^{\text{eff}} = 0.116W / m \cdotp K$ and $\tilde{\chi}^{\text{eff}} = 0$, and their imaginary parts are all zero. The numerical results demonstrate that for steady-state heat transfer (with periodic boundary conditions), the heat conduction in the composites is independent of the coupling terms, which agrees with the classic theory of heat conduction. 

When $\zeta = 0$, $\tilde{C}^{\text{eff}}$ in Fig. \ref{fig:zeta_0_unweight} (a) exhibits the clear trend that: the real part decreases monotonically with the normalized frequency, while its imaginary part increases from $0$ to a the peak at a low frequency around $\overline{\omega} = 0.082$, which reflects that the thermal performances is gradually shifted from quasi-static heat storage to a phase-lagged response. Compared to $\tilde{C}^{\text{eff}}$, $\tilde{K}^{\text{eff}}$ exhibits an opposite trend that: the real part increases with the normalized frequency, and its imaginary (magnitude) increases, which reveals the increasing dissipative contribution by the temperature gradient. Fig. \ref{fig:zeta_0_unweight} (c) plots the coupling term $\tilde{\chi}^\text{eff}$ ($\tilde{\chi}^{\text{eff}} = i \omega \tilde{\xi}^\text{eff}$). The coupling term is intensely frequency-dependent, so that the real part displays a peak at a low frequency around $\overline{\omega} = 0.082$ (same as $\tilde{C}^\text{eff}$ and then decreases. At the same time, its imaginary part decreases to a negative valley. The variation in coupling terms shows that coupling effects are not an artificially created concept but an intrinsic, frequency-dependent, and nonlocal effect caused by the microstructure. 

When $\zeta = 1$, the overall trends of effective properties remain the same, but some minor quantitative changes can be observed in Fig. \ref{fig:zeta_1_unweight}, which is consistent with the fact that $\zeta$ determines the macroscopic thermal wavenumber applied on the unit cell. For instance, real part of the peak value of $\tilde{C}^{\text{eff}}$ decreases from $1$ in Fig. 4(a) to $0.955$ in Fig. 5(a) at $\overline{\omega}=0$, and the peak value of its imaginary part decreases from $0.231$ to $0.194$, and the peak normalized frequency $\overline{\omega}$ shifts from $0.082$ to $0.062$; $\tilde{K}^\text{eff}$ varies with $\overline{\omega}$ monotonically in the opposite trends for the real and imaginary part, respectively, and using the values at $\overline{\omega}=0.625$ for example, its real part slightly increases from $0.216$ to $0.224$, and the imaginary part decreases from $0.147$ to $0.138$, respectively. The coupling term $\tilde{\chi}^\text{eff}$ exhibits the modest variation. For the real part, it decreases rapidly from zero at $\overline{\omega} = 0$ to a negative minimum around $-0.105$ at $\overline{\omega} \approx 0.27$, and increases to approximately $-0.096$ at $\overline{\omega} = 0.625$. For the imaginary part, it increases from zero to approximately $0.052$ at $\overline{\omega} \approx 0.062$, and then gradually decreases to $-0.029$ at $\overline{\omega} = 0.625$.
The above differences indicate that the extracted effective properties depend on both $\overline{\omega}$ and $\zeta$. Moreover, the excellent agreement among the three homogenization schemes indicates that the proposed frameworks can capture the coupled, frequency- and thermal-wavenumber-dependent effective thermal performance in a consistent manner.


\begin{figure}
    \centering
    \includegraphics[width=\linewidth]{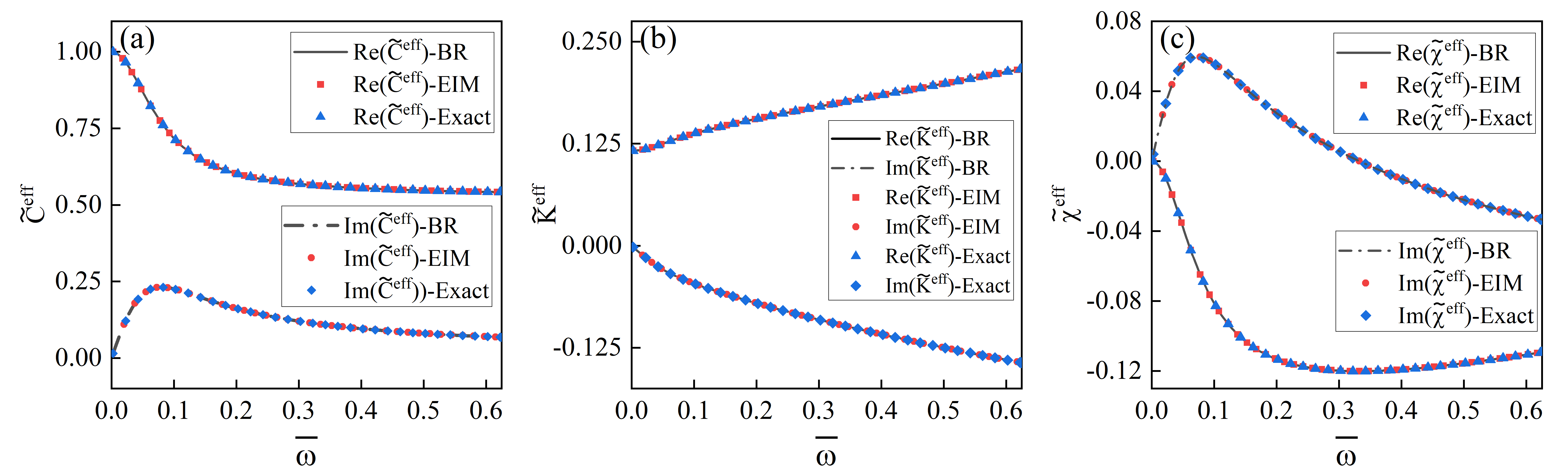}
    \caption{Comparison of three homogenization schemes with the normalized frequency $\overline{\omega} \in [0, 0.625]$, BR: boundary retrieval method, EIM: equivalent inclusion method, and Exact: exact homogenization with source-driven method. (a) Variation of $\tilde{C}^{\text{eff}}$; (b) variation of $\tilde{K}^{\text{eff}}$; and (c) variation of $\tilde{\chi}^{\text{eff}}$, when macroscopic thermal wavenumber $\zeta = 0$. }
    \label{fig:zeta_0_unweight}
\end{figure}


\begin{figure}
    \centering
    \includegraphics[width=\linewidth]{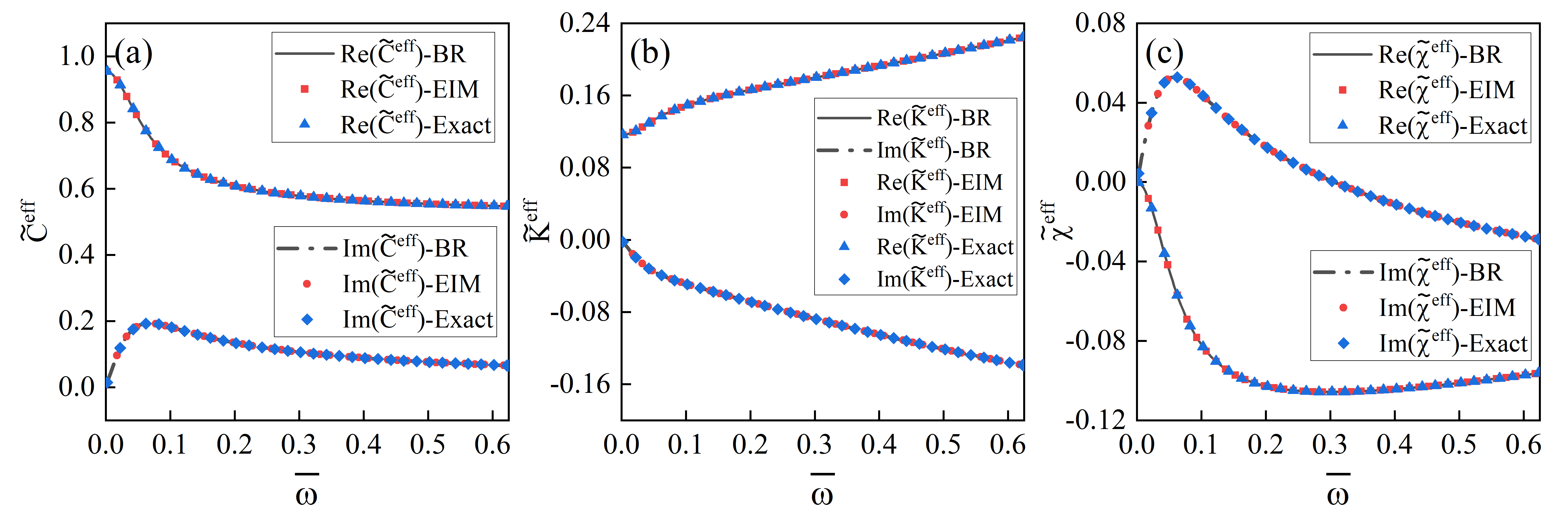}
    \caption{Comparison of three homogenization schemes with the normalized frequency $\overline{\omega} \in [0, 0.625]$, BR: boundary retrieval method, EIM: equivalent inclusion method, and Exact: exact homogenization with source-driven method. (a) Variation of $\tilde{C}^{\text{eff}}$; (b) variation of $\tilde{K}^{\text{eff}}$; and (c) variation of $\tilde{\chi}^{\text{eff}}$, when  macroscopic thermal wavenumber $\zeta = 1$. }
    \label{fig:zeta_1_unweight}
\end{figure}

\subsection{Comparison of homogenization schemes on weighted average}
\begin{figure}
    \centering
    \includegraphics[width=\linewidth]{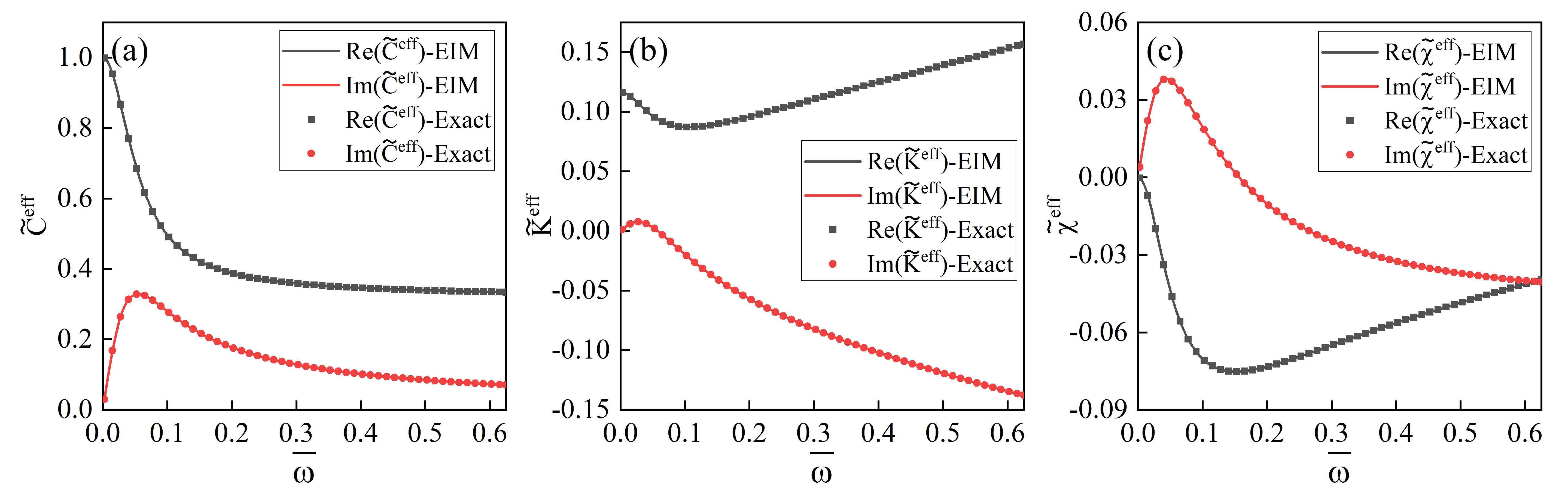}
    \caption{Comparison of two homogenization schemes with the normalized frequency $\overline{\omega} \in [0, 0.625]$, EIM: equivalent inclusion method, and Exact: exact homogenization with source-driven method. (a) Variation of $\tilde{C}^{\text{eff}}$; (b) variation of $\tilde{K}^{\text{eff}}$; and (c) variation of $\tilde{\chi}^{\text{eff}}$, when macroscopic thermal wavenumber $\zeta = 0$. The weight function is a piece-wise case, $f(x) \equiv \frac{\exp[-i \zeta x]}{c_2}$ in the second phase.}
    \label{fig:weight_zeta_0}
\end{figure}

\begin{figure}
    \centering
    \includegraphics[width=\linewidth]{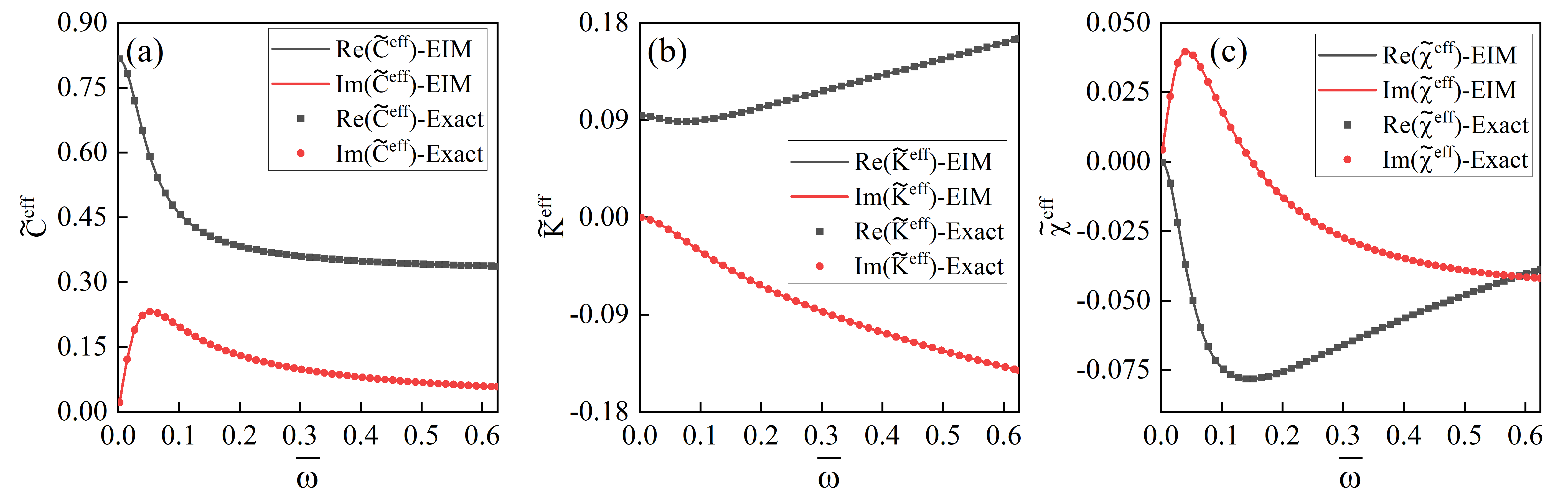}
    \caption{Comparison of two homogenization schemes with the normalized frequency $\overline{\omega} \in [0, 0.625]$, EIM: equivalent inclusion method, and Exact: exact homogenization with source-driven method. (a) Variation of $\tilde{C}^{\text{eff}}$; (b) variation of $\tilde{K}^{\text{eff}}$; and (c) variation of $\tilde{\chi}^{\text{eff}}$, when macroscopic thermal wavenumber $\zeta = 1$. The weight function is a piece-wise case, $f(x)\equiv \frac{\exp[-i \zeta x]}{c_2}$ in the second phase.}
    \label{fig:weight_zeta_1}
\end{figure}

As previously mentioned, since the BR method cannot account for the internal weighting operation, it cannot be applied to weighted-average analysis, so it is dropped off from the following comparison of the effect of weight function on the homogenization properties. Figs. \ref{fig:weight_zeta_0} and  \ref{fig:weight_zeta_1} show the extracted effective properties with a piecewise weight function, $f(x) = \exp[-i \zeta x] / c_2$ for the second phase, and zero elsewhere. With such a selection, the corresponding weighted ensemble-averaged Green's functions have been presented in Figs. \ref{fig:Gfunc_weighted} (a) and (b) with two normalized frequencies, and they should be interpreted in the sense of phase-filtered quantities, which emphasize the nonlocal properties of the second phase. Even under the intensive selectivity-weighted function, the EIM-implemented results have shown excellent agreement with those from the exact homogenization method for both real and imaginary parts, indicating that the two homogenization schemes and formulae are consistent. When the normalized frequency $\overline{\omega} = 0$, the effective properties in Fig. \ref{fig:weight_zeta_0} show a slight difference compared to Fig. \ref{fig:zeta_0_unweight}. For instance, the effective thermal conductivity changes from $0.11631$ to $0.11618$ (less than $0.01\%$).

Compared to Fig. \ref{fig:zeta_0_unweight} (a),  $\tilde{C}^\text{eff}$ in Fig. \ref{fig:weight_zeta_0} (a) exhibits a similar trend, where the real part monotonically decreases from $1$. and the imaginary part has a peak at a low frequency around $\overline{\omega} = 0.052$, which occurs earlier than in the unweighted case. In addition, the peak value increases from $0.231$ to $0.328$. Fig. \ref{fig:weight_zeta_0} (b) shows a different trend of $\tilde{K}^\text{eff}$ compared to Fig. \ref{fig:zeta_0_unweight} (b), where the real and imaginary parts do not monotonically change with the frequency, but exhibit a valley of the real part and a peak of the imaginary part with the extrema at different normalized frequencies, revealing a clear phase-lagging phenomenon caused by the filtering observation. The effective coupling terms $\tilde{\chi}^\text{eff}$ in Fig. \ref{fig:weight_zeta_0} (c) exhibit similar trends compared to Fig. \ref{fig:zeta_0_unweight} (c). The real part rapidly drops to a negative minimum and then gradually increases, while its imaginary part is positive at low normalized frequency, and becomes negative at higher normalized frequencies. Such phenomena, including the change of sign, phase shifting, and occurrences of extrema, are typical evidence of nonlocal coupling induced by the microstructural heterogeneity, which are augmented by the phase-selective weighting. 

When $\zeta = 1$, the overall trends of effective properties are similar, but some minor quantitative changes can be found. The peak of $Im(\tilde{C}^{\text{eff}})$ and $Im(\tilde{\chi}^\text{eff})$ are slightly decreased from $0.052$ to $0.05$. Compared Fig. \ref{fig:weight_zeta_1} (b) and Fig. \ref{fig:zeta_1_unweight} (b), the imaginary part of the effective property $\tilde{K}^\text{eff}$ decreases monotonically, which is different from the case $\zeta = 0$. In addition, greater discrepancies between the weighted and unweighted cases are observed in the magnitudes of the effective properties. For instance, the values at $\overline{\omega} = 0.625$ are listed and compared: (i) the $\tilde{C}^\text{eff}$ changes from $0.547 + 0.065 i$ to $0.334 + 0.071 i$; (ii) the $\tilde{K}^\text{eff}$ changes from $0.225 - 0.139 i$ to $0.157 - 0.138 i$; and (iii) the coupling term $\tilde{\chi}^{\text{eff}}$ changes from $-0.096 - 0.029 i$ to $-0.039 - 0.041 i$. Therefore, as previously discussed, the weight functions serve as tools for observation, and different effective thermal performances can be expected with different selections. 


\subsection{Comparison between the de-phased average and conventional average}
\begin{figure}
    \centering
    \includegraphics[width=\linewidth]{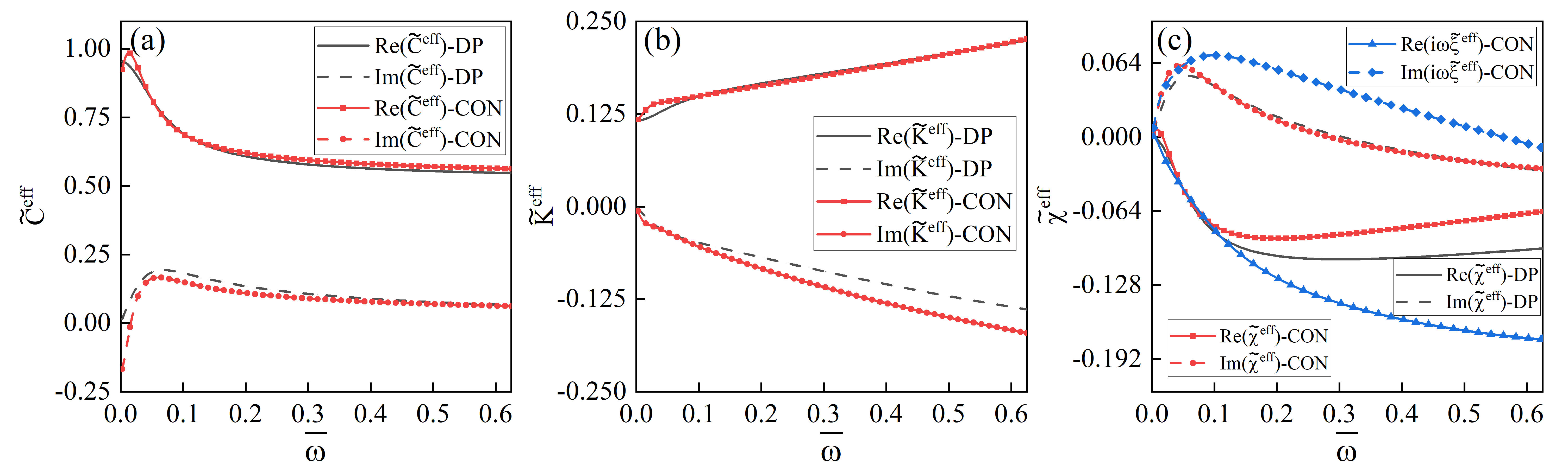}
    \caption{Comparison of two weight function in homogenization with the normalized frequency $\overline{\omega} \in [0, 0.625]$, DP: de-phased $f(x) = \exp[-i \zeta x]$, and CON: conventional $f(x) = 1$. (a) Variation of $\tilde{C}^{\text{eff}}$; (b) variation of $\tilde{K}^{\text{eff}}$; and (c) variation of $\tilde{\chi}^{\text{eff}}$ and $i \omega \tilde{\xi}^\text{eff}$, when  macroscopic thermal wavenumber $\zeta=1$. }
    \label{fig:weight_cmp}
\end{figure}

Under Bloch-form boundary conditions, the de-phased average employs $f(x) = \exp[-i \zeta x]$ required by Willis' concept, while the conventional average generally mixes multiple Fourier components and therefore may violate the coupling correlation. Although we have demonstrated that the exact homogenization, EIM-implemented, and the boundary-retrieval method agree well with each other for $f(x) = \exp[-i \zeta x]$, it is important to show the conventional case with weight function $f(x) = 1$, which is a common choice in steady-state analysis. In Fig. \ref{fig:weight_cmp}, (i) the de-phased averages are evaluated by the exact homogenization method, which is denoted as DP; and (ii) the conventional averages are evaluated by the EIM-implemented method, which is denoted as CON. 

When $\zeta = 1$, Figs. \ref{fig:weight_cmp} (a-c) present the variation of effective properties extracted using the de-phased average and conventional average weight functions, respectively. We can observe: (a) Effective property $\tilde{C}^\text{eff}_p$ from the DP and CON methods are very close, as the primary differences only exist in the low-frequency zone, which indicates that $\tilde{C}^\text{eff}$ is comparatively insensitive to the weight functions at $\zeta = 1$; (b) The two weight functions provides close predictions on $Re(\tilde{K}^{\text{eff}})$, but apparent discrepancies can be observed for the imaginary part with higher frequencies; (c) In contrast, the coupling terms are much more sensitive to the weight choices. Fig. \ref{fig:weight_cmp} (c) shows that the DP and CON results exhibit obvious discrepancies on $Im(\tilde{\chi}^{\text{eff}})$ starting from $\overline{\omega} = 0.1$. Moreover, using the de-phased average, the two coupling terms are correlated with the factor $i \omega$. However, the two blue curves on $i \omega \tilde{\xi}^{\text{eff}}$ do not match the two red curves on $\tilde{\xi}^\text{eff}$, which indicates that the coupling terms by the conventional average do not preserve such relations. Therefore, the conventional average fails to isolate the $\kappa = \zeta$ Bloch component and mixing multiple Fourier contributions. In contrast, the de-phased average performs well of the proper $\zeta$-projection, which is consistent with Willis' formulation.

\subsection{Effect of the asymmetry on thermal impedance and coupling terms}
In Section 3, Eq. (\ref{eq:dephase_shift}) shows that under the Bloch-form boundary condition, the microstructure with any realization variable $Y$ exhibits the same effective properties, including its expectation, the ensemble-averaged effective properties. When the concentrated heat capacity vanishes, the coupling terms $\tilde{\xi}^\text{eff}$ and $\tilde{\chi}^\text{eff}$ are zero, which was previously discussed in Section 5b. Therefore, the thermal coupling terms should arise from the microstructure's asymmetry, such as a concentrated heat capacity. 

\begin{figure}
    \centering
    \includegraphics[width=0.7\linewidth]{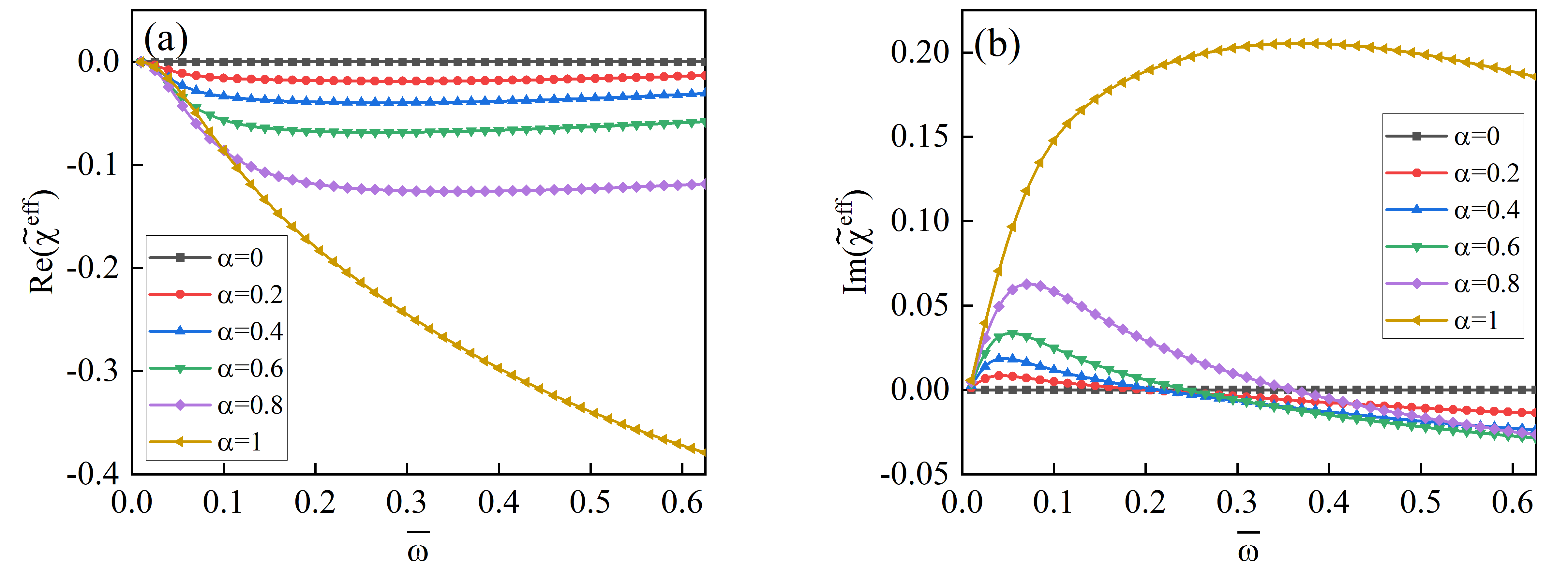}
    \caption{Comparison of the coupling term $\tilde{\chi}^\text{eff}$ in homogenization with the normalized frequency $\overline{\omega} \in [0, 0.625]$, when the position of the concentrated heat capacity $p = \alpha c_1 L$ ($\alpha = 0, 0.2, 0.4, 0.6, 0.8$ and $1$). (a) Real and (b) imaginary part of $\tilde{\chi}^\text{eff}$, when macroscopic thermal wavenumber $\zeta=1$.}
    \label{fig:alpha_chi}
\end{figure}

The following sets the realization variable $Y = 0$ and employs the de-phased average. To investigate the effects of the asymmetry, the concentrated heat capacity is placed at $p = \alpha c_1 L$, where $\alpha = 0, 0.2, 0.4, 0.6, 0.8$ and $1$. Figs. \ref{fig:alpha_chi} (a) and (b) present the real and imaginary parts of the coupling term $\tilde{\chi}^\text{eff}$ with the normalized frequency $\overline{\omega}$. When $\alpha = 0$, the concentrated heat capacity is placed at the center of the laminates, and thus the microstructure is symmetric, which exhibits zero coupling terms. However, when the concentrated heat capacity is not placed at the center, the coupling term occurs, and it increases with $\alpha$. For instance, at the normalized frequency $\omega = 0.625$, $\text{Re}(\tilde{\chi}^\text{eff})$ are $-0.013$, $-0.058$, and $-0.378$ for cases of $\alpha = 0.2, 0.6$ and $1$, respectively, which shows that, under the realization $Y=0$, a greater $\alpha$ enhances the extent of microstructure asymmetry, and therefore the magnitude of the coupling term significantly increases. In addition, Fig. \ref{fig:alpha_chi} (b) shows that at certain frequencies, the imaginary part of the coupling term may become zero. Since the coupling coefficient is complex-valued under harmonic heat transfer, the positive and negative imaginary parts only indicate the transition in the phase of the Willis coupling contribution, instead of the disappearance of the coupling effect.

\begin{figure}
    \centering
    \includegraphics[width=\linewidth]{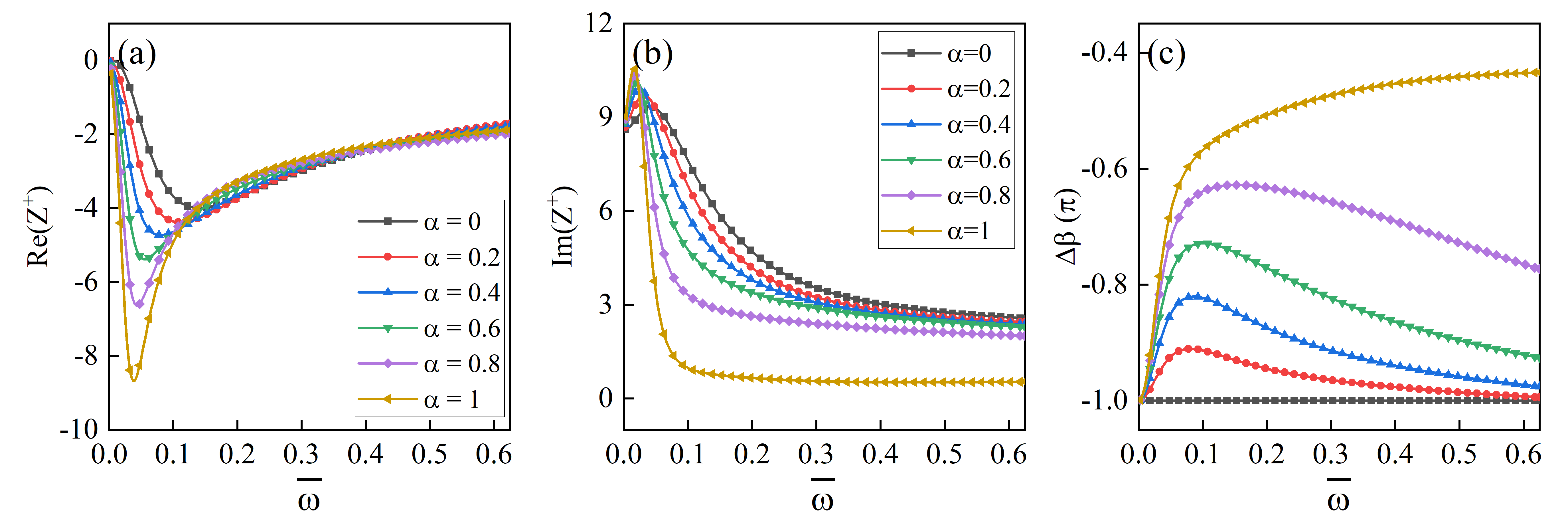}
    \caption{Comparison of the impedance $Z^+$ in homogenization with the normalized frequency $\overline{\omega} \in [0, 0.625]$, when the position of the concentrated heat capacity $p = \alpha c_1 L$ ($\alpha = 0, 0.2, 0.4, 0.6, 0.8$ and $1$). (a) Real and (b) imaginary part of $Z^+$; and (c) the difference between the phase angle $\Delta \beta$, when macroscopic thermal wavenumber $\zeta=1$.}
    \label{fig:imp}
\end{figure}

In addition to the thermal coupling term, Fig. \ref{fig:imp} further presents the variation of the thermal impedance $Z^+$ \cite{Gal2025a} and the phase angle between the opposite thermal waves, which are defined as, 

\begin{equation}
    Z^\pm = \frac{\langle T^\pm \rangle}{\langle q^\pm \rangle} = \frac{\langle T^\pm \rangle}{\chi \langle T^\pm \rangle \mp (i \zeta) \tilde{K}^\text{eff} \langle T^\pm \rangle} = \frac{1}{\chi \mp (i \zeta)\tilde{K}^\text{eff}} \quad \text{and} \quad \Delta \beta = \text{arg} \left( \frac{Z^+}{Z^-} \right)
    \label{eq:def}
\end{equation}
where $Z^\pm$ refers to the thermal impedance with the thermal wavenumber $\zeta$ and its opposite direction $-\zeta$, respectively; $\Delta \beta$ represents the phase angle difference between $Z^\pm$, in which $\text{arg}(.)$ evaluates the complex phase angle. Note that the classical Fourier law is modified by the additional averaged temperature-related coupling term $\tilde{\chi}^\text{eff}$. According to Figs. \ref{fig:imp} (a) and (b), when $\alpha = 0$, the thermal impedance is controlled by the $i \zeta \tilde{K}$, which follows the standard diffusive thermal behavior. As shown in Fig. \ref{fig:imp} (c), the phase angle difference is $-\pi$ ($\alpha = 0$), which implies that the heat transfer is symmetric. When $\alpha$ increases, the asymmetry of the microstructure introduces a nonzero coupling term, which changes both the magnitude and phase of the impedance, see the real and imaginary parts in Figs. \ref{fig:imp} (a) and  (b). For instance, the real part of the thermal impedance decreases at low frequencies with increasing $\alpha$, indicating a phase lag that differs from the standard diffusive behavior. The increasing magnitudes of the real and imaginary parts in the low-frequency region demonstrate that the coupling effect significantly redistributes the contributions from temperature and temperature gradients. Moreover, Fig. \ref{fig:imp} (c) shows that with increasing $\alpha$, the phase angle difference gradually deviates from the standard symmetric diffusive behavior, i.e., $\Delta \beta = -\pi$. This implies the Willis thermal coupling breaks the simple antisymmetric relation between two thermal impedances in opposite directions. Specifically, the phase relationship between the averaged temperature and heat flux is direction-dependent rather than a simple mirror projection. Therefore, the above results clearly reveal that the thermal Willis coupling is not only a formal extension of the constitutive relation, but affects observable macroscopic thermal quantities such as the direction-dependent thermal impedance. 

\section{Conclusions}
We have analyzed periodically layered conductors using three different homogenization methods, thereby providing the first explicit calculation of nonlocal thermal bianisotropy in a time-independent composite, where the coupling is induced by spatial asymmetry rather than by spatiotemporal modulation.

We treated the laminate as a random medium by considering different translations of the unit cell as different realizations, in which case ensemble-averaging is equal to averaging over the periodic part of the Bloch-Floquet form. Using the periodic Green's function of the laminate, we obtained explicit Fourier-space expressions for the effective conductivity, heat-capacity kernel, and the two thermal bianisotropic coupling kernels. These expressions show that the two coupling terms are correlated by the factor \(i\omega\), in agreement with the general thermal-bianisotropic theory.

We verified these exact expressions using two additional homogenization method. The first is a polarization formulation based on an infinite-domain Green's function and equivalent eigen-fields, which avoids relying on the explicit Green's function of the laminate. The second is a modified boundary-retrieval method. For the de-phased average, all three methods give consistent effective properties. This agreement confirms that the boundary-retrieval procedure captures the same thermal bianisotropic response as the averaging-based homogenization, provided that the macroscopic fields are defined by the appropriate Bloch projection.

The calculations also clarify the microstructural origin of the coupling in the effective representative considered here. The localized heat-capacity inclusion breaks the reflection symmetry of the unit cell. When the inclusion is placed at the center of the unit-cell, the bianisotropic kernels vanish; when it is displaced, the coupling terms become nonzero. Varying the inclusion position therefore provides a direct way to tune both the magnitude and phase of the thermal-bianisotropic response. This interpretation should be understood in light of the fact that no residual temperature-gradient field is introduced in the present formulation. If such a residual field were admitted, certain nonlocal effects would be represented by the cross-couplings rather than being absorbed into the direct effective kernels \cite{Sieck2017prb,PernasSalomon2020,Muhafra2022}.

A further implication is obtained from the effective thermal impedance. The bianisotropic kernels render the impedance direction dependent, changing the phase and magnitude of the macroscopic thermal response for opposite propagation directions. This directional impedance provides a physical signature of thermal bianisotropy and connects the formal nonlocal kernels to a measurable effect in asymmetric thermal metamaterials.

Finally, we showed that the averaging procedure is not a technical detail but part of the effective description. The de-phased average isolates the relevant Bloch component and preserves the Willis-type correlation between the coupling kernels. By contrast, the conventional spatial average mixes several Fourier components and can produce apparent effective properties that do not satisfy this correlation. This observation reinforces the importance of defining macroscopic fields consistently when extracting nonlocal effective properties.

The present work is restricted to one-dimensional periodic laminates, but the polarization formulation based on the infinite-domain Green's function is not restricted to this geometry. It therefore provides a route toward calculating thermal bianisotropic properties in more general two- and three-dimensional composites, where an explicit microstructure-specific Green's function is generally unavailable. Collectively, these results may expand the toolbox for thermal metamaterials.
\section*{CRediT Author Statement}
\textbf{Chunlin Wu}: Conceptualization, Methodology, Data Curation, Software, Validation, Writing-Original Draft, Funding Acquisition; \textbf{Gal Shmuel}: Writing-Review \& Editing; \textbf{Huiming Yin}: Conceptualization, Writing-Review \& Editing; Funding Acquisition. 

\section*{Acknowledgment}
CW's work is supported by the Chenguang Program No. 23CGA50 of Shanghai Education Development Foundation and Shanghai Municipal Education Commission, National Natural Science Foundation of China Grant No. 12302086. GS acknowledges funding by the U.S.~MURI project  W911NF-25-2-0176 and by the European Union (ERC, EXCEPTIONAL, Project No.~101045494). Funded by the European Union. The views and opinions expressed are those of the author(s) only and do not necessarily reflect those of the European Union or the European Research Council Executive Agency. Neither the European Union nor the granting authority can be held responsible for them. HY's work is sponsored by U.S. Department of Agriculture NIFA \#2021-67021-34201, and NIFA SBIR \#20233353039686, and National Science Foundation (NSF) grants IIP \#1738802 and IIP \#1941244. Those supports are gratefully acknowledged.

\appendix

\section*{Appendix A: Definition of Fourier coefficients for Green's function}

\subsection*{(i) Fourier coefficients $a_m(\mu)$ and $a_{-m}(-\mu)$}
In Section 4, the Fourier coefficients can be obtained as follows:
\begin{equation}
\begin{aligned}
    a_m(\mu) = \frac{1}{2L} \int_{-L}^{L} \psi_+(x) \exp \left[ \frac{i m \pi x}{L} \right] \thinspace dx = \frac{1}{2L} \int_{-L}^{L} \exp[-\mu x] \phi_+(x) \exp \left[ \frac{i m \pi x}{L} \right] \thinspace dx \\ 
    a_m(-\mu) = \frac{1}{2L} \int_{-L}^{L} \psi_-(x) \exp \left[ \frac{i m \pi x}{L} \right] \thinspace dx = \frac{1}{2L} \int_{-L}^{L} \exp[\mu x] \phi_-(x) \exp \left[ \frac{i m \pi x}{L} \right] \thinspace dx
\end{aligned}
\label{eq:a_coefficient}
\end{equation}
With all coefficients in Eq. (\ref{eq:form_sol}), the integral in Eq. (\ref{eq:a_coefficient}) can be derived explicitly. Specifically, it can be written as a superposition of three components, 

\begin{equation}
    a_m(\mu) = a_{m}^{\alpha}(\mu) + a_{m}^{1-\alpha}(\mu) + a_{m}^{L-c_1} (\mu)
    \label{eq:a_coef}
\end{equation}
where the detailed expression for $a_{m}^{\alpha}(\mu)$, $a_{m}^{1-\alpha}(\mu)$, and $a_{m}^{L-c_1} (\mu)$ are elaborated in the Supplemental Material 1.2. Note that the other coefficient $a_{m}(-\mu)$ shares a similar formula to Eq. (\ref{eq:a_coef}); however, all coefficients, $A, B, c, d$ should be reevaluated when the Floquet number is $-\mu$. Subsequently, two additional Fourier coefficients are defined as, 

\begin{equation}
\begin{aligned}
    a_m^w(\mu) = \frac{1}{2L} \int_{-L}^{L} w(x) \psi_+(x) \exp \left[ \frac{i m \pi x}{L} \right] \thinspace dx = \frac{1}{2L} \int_{-L}^{L} \exp[-\mu x] w(x) \phi_+(x) \exp \left[ \frac{i m \pi x}{L} \right] \thinspace dx \\ 
    a_m^w(-\mu) = \frac{1}{2L} \int_{-L}^{L} w(x) \psi_-(x) \exp \left[ \frac{i m \pi x}{L} \right] \thinspace dx = \frac{1}{2L} \int_{-L}^{L} \exp[\mu x] w(x) \phi_-(x) \exp \left[ \frac{i m \pi x}{L} \right] \thinspace dx
\end{aligned}
\label{eq:aw_coef}
\end{equation}
As previously mentioned, the weight function can be defined based on the specific use. When the weight function is $1$, the above equation reduces to Eq. (\ref{eq:a_coefficient}). The evaluation of Eq. (\ref{eq:aw_coef}) shares a similar procedure to Eq. (\ref{eq:a_coefficient}). For instance, if the weight function is only non-zero ($1 / c_2$) in the second phase, $a_m^w(\mu) = \frac{1}{c_2} a_m^{L-c_1}(\mu)$, and the only difference exists in the multiplication of the weight function. 

\subsection*{(ii) Fourier coefficients $b_m(\mu)$, $b_{-m}(-\mu)$, $c_m(\mu)$ and $c_{-m}(-\mu)$}
Similarly, the other Fourier coefficients can be obtained as follows:

\begin{equation}
    \begin{aligned}
    b_m(\mu) &= \frac{-1}{2L} \int_{-L}^{L} K_{Y=0}(x) \exp [-\mu x] \frac{d \phi_+(x)}{dx} \exp \left[ \frac{i m \pi x}{L} \right] \thinspace dx\\
    b_m(-\mu) &= \frac{-1}{2L} \int_{-L}^{L} K_{Y=0}(x) \exp [\mu x] \frac{d \phi_-(x)}{dx} \exp \left[ \frac{i m \pi x}{L} \right] \thinspace dx \\ 
    c_m(\mu) &= \frac{1}{2L} \int_{-L}^{L} C_{Y=0}(x) \exp [-\mu x] \phi_+(x) \exp \left[ \frac{i m \pi x}{L} \right] \thinspace dx\\
    c_m(-\mu) &= \frac{1}{2L} \int_{-L}^{L} C_{Y=0}(x) \exp [\mu x] \phi_-(x) \exp \left[ \frac{i m \pi x}{L} \right] \thinspace dx
    \end{aligned}
\end{equation}
Note that the four Fourier coefficient $b_m(\mu), c_m(\mu)$ and $b_{-m}(-\mu), c_{-m}(-\mu)$ are not fully independent. 
As the Green's function satisfies Eq. (\ref{eq:GreenPDE}), the components $\phi_+(x)$ and $\phi_-(x)$ should satisfy the following condition, 

\begin{equation}
    \frac{d \left( K_0(x) \frac{d\phi_{\pm}(x)}{dx}  \right)}{dx} = -i \omega C_{Y=0}(x) \phi_{\pm}(x)
    \label{eq:cond_phi}
\end{equation}
Therefore, $c_m(\mu)$  can be written in terms of partial derivatives of the heat flux, 

\begin{equation}
\begin{split}
    c_m(\mu) & = \frac{1}{i \omega}\int_{-L}^{L} \exp [-\mu x] \frac{d \left( -K_0(x) \frac{d\phi_{+}(x)}{dx}  \right)}{dx} \exp \left[ \frac{i m \pi x}{L} \right] \thinspace dx \\
    & = \frac{1}{i \omega} \int_{-L}^{L} \frac{d}{dx} \left\{ \exp \left[ -\mu x + \frac{i m \pi x}{L} \right] \left( -K_0(x) \frac{d \phi_+(x)}{dx} \right) \right\} dx \\ & - \frac{1}{i \omega} \left(-\mu + \frac{i m \pi}{L} \right) \int_{-L}^{L} -K_0(x) \frac{d\phi_{+}(x)}{dx} \exp \left[ -\mu x + \frac{i m \pi x}{L} \right] dx \\
    &= \frac{1}{i \omega} \left( \mu - \frac{i m \pi}{L} \right) b_m(\mu)
\end{split}
    \label{eq:c_coef_2}
\end{equation}
and similarly $c_{-m}(-\mu)$ and $b_{-m}(-\mu)$ satisfy: 
\begin{equation}
    c_{-m}(-\mu) = - \frac{1}{i \omega} \left( \mu - \frac{i m \pi}{L} \right) b_{-m}(-\mu)
    \label{eq:c_coef_2m}
\end{equation}
Eqs. (\ref{eq:c_coef_2}) and (\ref{eq:c_coef_2m}) will be subsequently applied in the extraction of effective properties in the following subsection. With all coefficients in Eq. (\ref{eq:form_sol}), four coefficients can be explicitly expressed as, 

\begin{equation}
    b_{m}(\mu) = b_{m}^{(\alpha)}(\mu) + b_{m}^{(1 - \alpha)}(\mu)+ b_{m}^{(L - c_1)}(\mu)
\end{equation}
where $b_{m}^{(\alpha)}(\mu)$, $b_{m}^{(1 - \alpha)}(\mu)$, and $b_{m}^{(L - c_1)}(\mu)$ are elaborated in the Supplemental Material 1.2. Finally, the $c_m(\mu)$ coefficients can be determined as: 

\begin{equation}
\begin{split}
    c_m(\mu) = &C^1 \left[a_m^{(\alpha)}(\mu) + a_m^{(1 - \alpha)}(\mu) \right] + C^2 a_m^{(L - c_1)}(\mu) \\
    &+ \frac{H\left(\cosh [k_1 \alpha c_1 L] + b \sinh [k_1 \alpha c_1 L] \right)}{2L} \exp\left[ \alpha c_1 (i m \pi - \mu L) \right]
\end{split}
\end{equation}

\end{document}